\documentclass[final,hidelinks,onefignum,onetabnum]{siamart250211}

\usepackage{lipsum}
\usepackage{amsfonts}
\usepackage{epstopdf}
\ifpdf
\DeclareGraphicsExtensions{.eps,.pdf,.png,.jpg}
\else
\DeclareGraphicsExtensions{.eps}
\fi

\usepackage{microtype}
\usepackage[scr=rsfs]{mathalpha}
\usepackage[english]{babel}

\usepackage[
textwidth=15cm,
textheight=24cm,
centering
]{geometry}

\usepackage{silence}

\usepackage{mathtools}
\usepackage{amssymb}
\usepackage{dsfont}
\usepackage{siunitx}
\usepackage{esint}

\numberwithin{equation}{section}
\allowdisplaybreaks

\usepackage{booktabs}
\usepackage{multirow}
\usepackage{graphicx}
\usepackage{xcolor}
\usepackage[caption=false]{subfig}

\usepackage{csquotes}

\usepackage{comment}

\DeclareMathOperator{\sign}{sign}
\DeclareMathOperator{\Tr}{Tr}
\DeclareMathOperator{\TrLim}{\underline{Tr}}
\DeclareMathOperator{\Supp}{Supp}
\renewcommand{\Im}{\operatorname{Im}}
\renewcommand{\Re}{\operatorname{Re}}

\providecommand\given{}
\newcommand\SetSymbol[1][]{%
	\nonscript\:#1\vert
	\allowbreak
	\nonscript\:
	\mathopen{}}
\DeclarePairedDelimiterX\Set[1]\{\}{%
	\renewcommand\given{\SetSymbol[\delimsize]}
	#1
}

\newcommand{\N}{\mathbb{N}}
\newcommand{\Z}{\mathbb{Z}}
\newcommand{\Q}{\mathbb{Q}}
\newcommand{\R}{\mathbb{R}}
\newcommand{\C}{\mathbb{C}}

\newcommand{\PP}{\mathbb{P}}

\newcommand{\cH}{\mathcal{H}}
\newcommand{\cN}{\mathcal{N}}

\newcommand{\cS}{\mathcal{S}}

\newcommand{\cX}{\mathcal{X}}

\newcommand{\cK}{\mathcal{K}}

\newcommand{\cU}{\mathcal{U}}

\newcommand{\bG}{\mathbf{G}}

\newcommand{\bzero}{\mathbf{0}}

\newcommand{\unfold}{\mathcal{T}}
\newcommand{\WL}{\mathcal{D}_{W,L}}
\newcommand{\WLDOS}{\mathcal{D}_{WLh}}

\newsiamremark{remark}{Remark}
\newsiamremark{hypothesis}{Hypothesis}
\crefname{hypothesis}{Hypothesis}{Hypotheses}
\newsiamthm{claim}{Claim}
\newsiamremark{fact}{Fact}
\crefname{fact}{Fact}{Facts}
\newsiamthm{assumption}{Assumption}
\crefname{assumption}{Assumption}{Assumptions}
\crefname{ALC@unique}{Line}{Lines}
\crefname{remark}{remark}{remarks}
\Crefname{remark}{Remark}{Remarks}
\AddToHook{env/remark/begin}{\crefalias{theorem}{remark}}
\crefname{appendix}{appendix}{appendices}
\Crefname{appendix}{Appendix}{Appendices}

\headers{DoS of multiscale Schr\"odinger operators}{E. Canc\`es, D. Massatt, L. Meng, \'E. Polack, and X. Quan}

\title{Numerical computation of the density of states of aperiodic multiscale Schr\"odinger operators}

\author{
	Eric Canc\`es
	\thanks{Eric Canc\`es,
		CERMICS,
		\'Ecole des Ponts, Institut Polytechnique de Paris, and Inria,
		6 and 8 av. Pascal,
		77455 Marne-la-Vall\'ee,
		France (\email{eric.cances@enpc.fr})}
	\and
	Daniel Massatt
	\thanks{Daniel Massatt,
		New Jersey Institute of Technology,
		NJ,
		USA (\email{daniel.massatt@njit.edu})}
	\and
	Long Meng
	\thanks{Long Meng,
		Center for Interdisciplinary Applied Mathematics \& Institute of Fundamental and Transdiciplinary Research,
		Zhejiang University,
		China (\email{longmeng@zju.edu.cn})}
	\and
	\'Etienne Polack
	\thanks{\'Etienne Polack,
		CERMICS,
		\'Ecole des Ponts ParisTech and Inria Paris,
		6 \& 8 av. Pascal,
		77455 Marne-la-Vall\'ee,
		France
		and
		Universit\'e Grenoble Alpes,
		CEA Grenoble,
		IRIG, MEM, NRX,
		17 av. des Martyrs,
		Grenoble F-38000,
		France (\email{etienne.polack@math.cnrs.fr})}
	\and
	Xue Quan
	\thanks{Xue Quan,
		Academy of Mathematics and Systems Science,
		Chinese Academy of Sciences,
		China (\email{xuequan@amss.ac.cn})}
}

\usepackage{amsopn}

\begin{document}

\maketitle

\begin{abstract}
 Computing the electronic structure of incommensurate materials is a central challenge in condensed matter physics, requiring efficient ways to approximate spectral quantities such as the density of states (DoS).
 In this paper, we numerically investigate two distinct approaches for approximating the DoS of incommensurate Hamiltonians for small values of the incommensurability parameters $\epsilon$ (e.g., small twist angle, or small lattice mismatch): the first employs a momentum-space decomposition, and the second exploits a semiclassical expansion with respect to $\epsilon$.
 In particular, we compare these two methods using a 1D toy model. We check their consistency by comparing the asymptotic expansion terms of the DoS, and it is shown that, for full DoS, the two methods exhibit good agreement in the small $\epsilon$ limit, while discrepancies arise for less small $\epsilon$, which indicates the importance of higher-order corrections in the semiclassical method for such regimes. We find these discrepancies to be caused by oscillations in the DoS at the semiclassical analogues of Van Hove singularities, which can be explained qualitatively, and quantitatively for $\epsilon$ small enough, by a semiclassical approach.
\end{abstract}



\section{Introduction}

Computing the electronic structure of crystalline materials and amorphous materials is one of the major challenges of condensed matter physics. Here in particular, we focus on moiré materials, which have become of great scientific interest since the discovery of unconventional superconductivity in twisted bilayer graphene at ``magic'' twist angle $\theta \approx 1.1^\circ$ in 2018~\cite{pablo2018}, and have become a platform for studying other exotic many-body effects, including correlated insulation and the fractional quantum Hall effect \cite{ledwith2020, bernevig2020}. These systems, however, have numerous electrons, often modeled as infinite, and due to competing periodicities there is no global periodicity, which prohibits the use of classical Bloch theory techniques. Two main difficulties have to be addressed in such quantum computations: first, such systems contain a macroscopic number of electrons ($\sim 3 \times 10^{23}$ electrons in one gram of carbon 12), and second, electrons interact via long range Coulomb interactions.
In this work we will address the first difficulty for moiré materials, which are typically modeled as aperiodic systems with an infinite number of electrons, by comparing two computational methods and models for approximating moiré materials under the single-particle approximation.

\medskip

In such single-particle approximations of the many-body Schr\"{o}dinger model, non-interacting ``quasi-electrons'' are subjected to an effective potential $V_{\rm eff}$ modeling the interactions with the nuclei and the other electrons.  From a mathematical point of view, this amounts to considering Hamiltonians of the form
\begin{equation} \label{eq:effective_Hamiltonians}
H = - \frac 12 \Delta + V_{\rm eff} \quad \text{acting on $L^2(\R^d)$}.
\end{equation}
In the real world, $d=3$, but the cases when $d=1$ and $d=2$ are interesting to test numerical methods, and also because some reduced models for polymers, thin films, or 2D materials are set in dimensions 1 or 2. Variants of this model include spin degrees of freedom (magnetization), external electromagnetic fields (computation of response functions such as the electrical conductivity), or internal magnetic potentials (quantum anomalous Hall effect). The effective potential can be purely empirical, or be obtained self-consistently as in Kohn--Sham Density-Functional Theory (KS-DFT). We assume from now on that $V_{\rm eff} \in C^\infty(\R^d) \cap L^\infty(\R^d)$, an assumption allowing one to deal with most practical cases. Under this assumption, the linear operator \cref{eq:effective_Hamiltonians} is essentially self-adjoint and bounded from below. The unique self-adjoint extension of $H$, still denoted by $H$, has domain $H^2(\R^d)$ and form domain $H^1(\R^3)$. The nature of $\sigma(H)$, the spectrum of $H$, strongly depends of the properties of $V_{\rm eff}$.

\medskip

The electronic ground-state density matrix of the system is formally defined by
\begin{align} \label{eq:GSDM}
&\gamma_0 = \mathds 1_{(-\infty,\mu_{\rm F}]}(H), \\
& 2 \underline{\Tr}(\gamma_0)=N, \label{eq:GSDM2}
\end{align}
where $\TrLim$ is the trace per unit volume, i.e.,
\begin{equation}\label{eq:trace_per_unit_volume}
\TrLim(\gamma_0) \coloneq \lim_{R \to +\infty} \frac{1}{(2R)^d} \Tr\left( \chi_{[-R,R]^d} \gamma_0 \chi_{[-R,R]^d} \right),
\end{equation}
$N$ the number of electrons per unit volume in the system, and
$\mu_{\rm F}$ the Fermi level, that is the chemical potential associated with the constraint $2 \TrLim(\gamma_0)=N$ (the factor $2$ comes from the spin). The electronic ground-state density is the unique function $\rho_0 \in L^1_{\rm loc}(\R^d)$ such that
\begin{equation} \label{eq:GSD}
\forall W \in L^\infty_{\rm c}(\R^d), \quad \int_{\R^d} \rho_0 W=2\Tr( \gamma_0 W),
\end{equation}
where $L^\infty_{\rm c}(\R^d)$ is the space of compactly supported essentially bounded functions. In the case of nonlinear KS-DFT, $V_{\rm eff}$ is a function of $\rho_0$. Formally, $\rho_0(x)=2\gamma_0(x,x)$, where $\gamma_0(x,x')$ is the integral kernel of $\gamma_0$. Lastly, the density of states (DoS) $\nu_H$ is formally defined as the positive measure such that
\begin{equation}\label{eq:DOS}
\forall f \in C^\infty_{\rm c}(\R), \quad \TrLim(f(H)) = \int_{\R} f(\epsilon) \, d\nu_H(\epsilon).
\end{equation}
Of course, it is not clear \textit{a priori} if these equations make sense for any $V_{\rm eff}$, and as a matter of fact, they do not. We need to assume some uniformity at the macroscopic scale to ensure that the limit in the definition~\cref{eq:trace_per_unit_volume} of the trace per unit volume actually exists. This is the case in two important settings
\begin{itemize}
\item the periodic setting describing perfect crystals,
\item the ergodic setting describing macroscopically homogeneous disordered systems, such as disordered crystals (doped semiconductors, alloys) or glassy materials, as well as incommensurate systems such as quasicrystals or  moir\'e materials.
\end{itemize}

\medskip

Recall that if $V_{\rm eff}$ is periodic, Bloch theory provides an efficient practical way to analyze the spectral properties of $H$, and to prove in particular that $\sigma(H)$ is purely absolutely continuous, hence $\nu_H \in L^1_{\rm loc}(\R)$. It also allows one to compute numerical approximations of the spectral decomposition of $\sigma(H)$, from which all the electronic properties of the crystal can be inferred. The theory easily extends to mean-field models of Hartree--Fock or KS-DFT types (see \cite{Hartreecrystals} and  for its relativistic analog \cite{Diraccrystals}). Numerous electronic structure simulation codes are able to compute KS-DFT band diagram. A systematic comparison of the accuracies and performances of these codes as in 2016 was done in the review article~\cite{reviewDFTcodes}. Let us also mention the recent DFTK package~\cite{DFTK}, written in Julia, in which we have implemented the semiclassical expansion method presented below. DFTK provides to the community a useful platform to develop new numerical methods for KS-DFT as well as other linear or nonlinear Schr\"odinger-type models (e.g., Gross--Pitaevskii equations) at a low entrance cost compared to other existing software. 

\medskip

The ergodic case is more involved, and at this point we restrict the discussion to the linear Schr\"{o}dinger model. The idea is to consider the specific Hamiltonian $H$ in \cref{eq:effective_Hamiltonians} as a generic element of a family of ergodic Schr\"odinger operators $(H_\omega)_{\omega \in \Omega}$, where $(\Omega,\mathcal T,\PP)$ is a probability space.
The operators $H_\omega$ are assumed to satisfy the covariance relation
$$
\forall R \in \mathbb{G}, \quad H_{\tau_R\omega} = U_R^* H_\omega U_R,
$$
where $\mathbb{G} \subset \R^d$ is a group of translation vectors, $U_R$ the translation operator on $L^2(\R^d)$ defined by $(U_R\phi)(x)=\phi(x-R)$, and $\tau$ the action of $\mathbb{G}$ on $\Omega$. In other words, $H_{\tau_R\omega}$ is unitary equivalent to $H_\omega$ and the former is obtained from the latter by simply shifting by $R$ the origin of the Cartesian frame. The key assumption is that the action $\tau$ is probability preserving and ergodic, in the sense that
\begin{itemize}
\item for all $A\in\mathcal T$ and $R \in \mathbb{G}$, we have $\PP(\tau_R(A))=\PP(A)$,
\item if for some $A \in \mathcal T$, we have $\tau_R(A)=A$ for all $R \in \mathbb{G}$, then $\PP(A) \in \{0,1\}$.
\end{itemize}
Under some technical assumptions, it follows from Birkhoff theorem that for almost all $\omega \in \Omega$, the spectrum and the density of states of $H_\omega$ are well-defined and are independent of $\omega$.

\medskip

As a matter of example, disordered crystals can be modeled by ergodic random Schr\"odinger operators, among which the continuous Anderson model on the square lattice, which reads
\begin{equation}\label{eq:ESO}
H_\omega = -\frac 12 \Delta + V_\omega \quad \text{with} \quad
V_\omega(x) = \sum_{k \in \Z^d} q_k(\omega)\, v(x-k),
\end{equation}
where $v \in C^\infty_{\rm c}(\R^d)$ s.t. $\int_{\R^d} v =-1$, and $(q_k)_{k \in \Z^d}$ are i.i.d.\ random variables on $(\Omega,\mathcal T,\PP)$ such that $q_k(\tau_R\omega)=q_{k+R}(\omega) > 0$ for all $R \in \Z^d$ and a.a. $\omega \in \Omega$. The study of the spectral properties of random Schr\"odinger operators is a very active research topic. We refer the interested reader to~\cite{Randomoperators} and references therein.

\medskip

We focus in this work on comparing two algorithms for computing DoS for incommensurate bilayer systems, a momentum-space method and a semiclassical expansion method \cite{CancesLong,planewave2025}. For the sake of simplicity, we restrict ourselves to Hamiltonians of the form
\begin{equation}
    \label{eq:Hepsilon}
    H_\epsilon = - \frac{1}{2} \Delta + V(x,(1+\epsilon) x) \quad \text{on $L^2(\R^d)$},
\end{equation}
where $V \in C^\infty(\R^d \times \R^d;\R)$ is $\mathbb Z^d$-periodic in each variable and $\epsilon \in \R_+ \setminus \Q_+$, but both methods apply to more complex settings such as effective atomic-scale models for twisted bilayer graphene. The Hamiltonian $H_\epsilon$ can be embedded in the family of ergodic Schr\"odinger operators
$$
H_{\epsilon,\omega} = - \frac{1}{2} \Delta + V(x,(1+\epsilon) x+\omega), \quad \omega \in \Omega \coloneq \mathbb T^d,
$$
where $\mathbb T^d \equiv \R^d/\Z^d$ denotes the $d$-dimensional torus. Indeed, endowing $\Omega$ with the usual Lebesgue measure on the torus, and introducing the ergodic action $\tau$ of the translation group $\Z^d$ on $\Omega$ defined by $\tau_R\omega = \omega + \epsilon R$, we have that for all $R \in \Z^d$,
\begin{align*}
H_{\epsilon,\tau_R\omega} &= - \frac{1}{2} \Delta + V(x,(1+\epsilon) x+ \tau_R \omega) = - \frac{1}{2} \Delta + V(x,(1+\epsilon) x+  \omega + \epsilon R) \\
&=  -\frac{1}{2} \Delta + V(x,(1+\epsilon) (x+R)+  \omega) = U_R^*H_{\epsilon,\omega} U_R,
\end{align*}
where we have used that $V(x,(1+\epsilon) (R+x)+  \omega) =  V(x,(1+\epsilon) x+  \omega+\epsilon R)$ as $V$ is $\Z^d$-periodic in each variable. This formalism allows one to prove that operators of the form \cref{eq:Hepsilon} with $\epsilon \in \R_+ \setminus \Q_+$ have a well-defined DoS. We restrict our focus to the numerical computation of the DoS $\nu_{H_\epsilon}$ for the incommensurate Hamiltonian \cref{eq:Hepsilon}, employing momentum-space and semiclassical methods for small positive values of $\epsilon$.

The momentum-space method exploits the smoothness of the potential $V(x_1,x_2)$ lifted into the higher dimensional space by using a momentum-space decomposition
\begin{equation}
\TrLim(f(H_\epsilon)) = \int_{\R^d} \langle \xi| f(H_\epsilon)|\xi\rangle\,d\xi
\end{equation}
where $|\xi\rangle$ represents the momentum basis {(planewaves, with proper normalizations)}.
In particular, each momentum $\xi$ couples to a discrete dense collection of momenta $\{\xi + 2\pi n + 2\pi (1+\epsilon)m\}_{n,m \in \Z^d}$ described by a local lattice model $\widehat{H}_\epsilon(\xi)$. This collection of lattice models is efficiently truncated to tune accuracy exploiting confinement from the kinetic energy $-\frac{1}{2}\Delta$ and Combes--Thomas estimates on the resolvent. A Chebyshev expansion is used to compute each $\langle \xi| f(H_\epsilon)|\xi\rangle$. This method has precise error bounds and computational cost estimates depending on truncation choices and integral discretization.

\medskip

The semiclassical expansion method is based on an asymptotic expansion of $\nu_{H_\epsilon}$ in powers of $\epsilon$, in the following sense
\begin{equation}\label{eq:expansion_DOS}
\TrLim(f(H_\epsilon)) = L_0(f) + \epsilon L_1(f) + \epsilon^2 L_2(f) + \cdots,
\end{equation}
where the $L_j$'s are well-defined distributions on $\R$.
The above asymptotic expansion was proved in~\cite{CancesLong} for a model of twisted bilayer graphene in energy ranges around the charge neutrality Fermi level, and the arguments can be easily extended to the simplest model under consideration here. This approach was inspired by works by Dimassi~\cite{dimassi1993developpements} and Panati--Spohn--Teufel~\cite{panati2003effective}, based on many previous works by various authors going back to Balezard--Konlein~\cite{Balezard-Konlein_85}, and uses tools from pseudodifferential calculus with operator-valued symbols.

\medskip

This article is organized as follows. In \Cref{sec:PW_method,sec:semiclassical_expansion}, we review the mathematical foundations of the momentum-space method and the semiclassical expansion method, respectively. In \Cref{sec:discretization}, we detail the discretization scheme for each method. In \Cref{sec:DOS}, we check the consistency of the two approaches by comparing the zero, first, and second-order terms of the semiclassical expansion with the limiting value at $\epsilon=0$ of the DoS $\nu_{H_\epsilon}$ (obtained by the momentum-space method), and its first and second-derivatives (estimated by finite differences). In \Cref{sec:effective_moire_scale_Hamiltonians}, we illustrate numerically the fact that near band edges, harmonic approximations derived from the semiclassical framework lead to reasonable effective Hamiltonians correctly describing the inverse moiré scale Van Hove singularities in the DoS for $\epsilon$ small enough. More precisely,
the operator $H_\epsilon$ in~\cref{eq:Hepsilon} can be seen as the Weyl quantization of the operator-valued symbol
$$
\R^d \times \R^d \ni (k,X) \mapsto h(k,X) \coloneq
\frac 12 (-i\nabla_x + k)^2 + V(x,X) \quad \text{on $L^2(\mathbb T^d)$}.
$$
For each $(k,X)$, the operator $h(k,X)$ is self-adjoint, bounded below, and has a compact resolvent, so that its spectrum is purely discrete and the sequence $(E_n(k,X))_{n \ge 1}$ of its eigenvalues (counting multiplicities) forms a non-decreasing sequence going to $+\infty$:
$$
\R^d \times \R^d \ni (k,X) \mapsto \sigma(h(k,X)) = \{E_j(k,X)\}_{j \ge 1} \subset \R.
$$
We further show that for $\epsilon$ small enough, a local study of the symbols $h(k,X)$ around the critical points $p=(k_0,X_0)$ of the function $E_j$ provides useful information on the oscillations of the DoS of $H_\epsilon$ in a small energy window close to $E_j(p)$. The proof of \Cref{th:2.4}, which provides computable formulae for $L_0(f)$, $L_1(f)$ and $L_2(f)$ as integrals over $(2\pi \mathbb T^d) \times \mathbb T^d$ of integrands depending linearly on $f$ and easily computable from the spectral decomposition of $h(k,X)$, is postponed until \Cref{sec:Appendix}.

\section{Model and numerical methods}
In this section, we review the mathematical foundations of the momentum-space method and the semiclassical method.
\subsection{Momentum-space method}
\label{sec:PW_method}

In the momentum-space framework, the objective is to describe $H_\epsilon$ through coupling of the momentum-space basis $e^{i \xi \cdot x}$. We first build some notation. We use the Fourier transform
\[
\mathcal{F}\psi(\xi) = \hat \psi(\xi) =  \int_{\mathbb{R}^d} e^{-i\xi \cdot x} \psi(x)\,dx,
\]
and define the dual lattices
\[
\mathbb L^* = 2\pi \mathbb{Z}^d.
\]
We denote the Fourier modes of the potential by
\begin{equation}
    \hat V_{{\bG}} = \int_{{[0,1)^d\times [0,1)^d}} V(x_1,x_2)e^{-i (G_1\cdot x_1 + G_2\cdot x_2)} \,dx_1\,dx_2, \quad {\bG} = (G_1,G_2) \in \mathbb L^* \times \mathbb L^*.
\end{equation}
Then,
\begin{equation}
    V(x,(1+\epsilon)x) = \sum_{ {\bG} \in \mathbb L^* \times \mathbb L^*} \hat V_{\bG} e^{i(G_1\cdot x + (1+\epsilon)G_2\cdot x)}.
\end{equation}
{
We will assume that the potential $V$ is an analytic function, which yields an exponential decay of the Fourier coefficients $\hat{V}_G$ with respect to $|G_1| + |G_2|$. This analyticity is a stronger condition than the previously stated global standing assumption that $V\in C^\infty$. This is made for the sake of clarity and simplicity in the presentation, ensuring that the plane wave approximations \eqref{discrete_pw_dos} exhibit a clear exponential convergence rate in \Cref{thm:error:ms}. Notably, the $C^\infty$ regularity of $V$ will lead to a super-algebraic convergence rate in \Cref{thm:error:ms}.
}
We illustrate the momentum basis coupling succinctly by taking the Fourier transform of $H_\epsilon \psi$ for $\psi$ in the Schwartz class 
\begin{align}\label{eq:Schwartz}
     \cS(\R^n;\C) \coloneq \Set*{ \phi \in C^\infty(\R^n,\C) \given \forall j \in \N, \; \cN_j(\phi) \coloneq \max_{|\alpha| \le j} \max_{|\beta| \le j} \max_{y \in \R^n} \left| y^{\alpha} \partial^{\beta}\phi(y) \right| < \infty }.
\end{align}
We have
\begin{equation}
    \begin{split}
  \forall \psi \in \cS(\R^n;\C), \quad  \mathcal{F} [H_\epsilon \psi](\xi) &=  \frac{1}{2} |\xi|^2\hat \psi(\xi) + \sum_{\bG \in \mathbb L^*\times\mathbb L^*} \int_{\mathbb R^d} \hat V_{\bG} e^{i(G_1+(1+\epsilon)G_2-\xi)\cdot x} \psi(x) \,dx \\
    &=\frac{1}{2} |\xi|^2 \hat \psi(\xi) + \sum_{\bG \in \mathbb L^*\times\mathbb L^*} \hat V_{\bG} \hat \psi(\xi - G_1 -(1+\epsilon)G_2).
    \end{split}
\end{equation}
Hence the wavevector $\xi$ has scattering channels to the wavevectors $\xi + G_1 + (1+\epsilon)G_2$ through the coefficients given by $\delta_{0G_1}\delta_{0G_2} \frac{1}{2}|\xi|^2 + \hat V_{\bG}$. Likewise, any $\xi + G_1 +(1+\epsilon)G_2$ has a channel to wavevector $\xi + G_1' + (1+\epsilon)G_2'$ for $G_1,G_1',G_2,G_2' \in \mathbb L^*$. The magnitudes of all these channels are described by a family of discrete operators. To this end, we define an unfolding map $\unfold_\xi : \cS(\R^n;\C) \rightarrow \mathbb{C}^{\mathbb L^* \times \mathbb L^*}$ by
\begin{equation}
    (\unfold_\xi \hat \psi)_{\bG} = \hat \psi(\xi + G_1 +(1+\epsilon)G_2), \quad \bG\in \mathbb L^* \times \mathbb L^*.
\end{equation}
We let $\widehat{H}_\epsilon(\xi) : \mathcal{D} \subset \ell^2(\mathbb L^* \times \mathbb L^*) \rightarrow \ell^2(\mathbb L^* \times \mathbb L^*)$ be defined by
\begin{equation}
    [\widehat{H}_\epsilon(\xi)]_{ \bG,\bG'} = \frac{1}{2}|\xi +G_1+(1+\epsilon)G_2|^2\delta_{\bG, \bG'} + \hat V_{G_1-G_1',G_2-G_2'}
\end{equation}
for $\bG, \bG' \in \mathbb L^* \times \mathbb L^*$. Next, we define the class of test functions used:
\begin{equation}\label{def:Lambda-zeta-delta}
\Lambda_{\zeta,\delta} \coloneq \Set*{ f \in C^\infty(\mathbb{R}) \given
    \begin{aligned}
        & f \text{ admits an analytic extension to } z \text{ satisfying } |\Im(z)| < \delta, \\
        & \text{and } |f(z)| \leq C e^{-\zeta |{\rm Re}(z)|}, \; C >0
    \end{aligned}
}
\end{equation}
for $\delta,\zeta > 0$. We note that this class includes Gaussian functions, which are used in this work. In this Gaussian case, the optimal choice of $\delta$ corresponds to the standard deviation of the Gaussian, and hence to the spectral accuracy of the density of states. The smaller the standard deviation of the Gaussian test function, the finer the spectral information obtained. The following results hold by Lemma B.1 in \cite{planewave2025}:
\begin{proposition}
    For $\psi \in \cS(\R^n;\C)$, $\delta,\zeta > 0$, we have for $f \in \Lambda_{\zeta,\delta}$,
    \begin{align*}
        & \unfold_\xi [\mathcal{F}H_\epsilon\psi] = \widehat{H}_\epsilon(\xi) \unfold_\xi \mathcal{F}\psi \\
        & \unfold_\xi [\mathcal{F}f(H_\epsilon)\psi] = f(\widehat{H}_\epsilon(\xi)) \unfold_\xi \mathcal{F} \psi
    \end{align*}
\end{proposition}

In particular, $[f(\widehat{H}_\epsilon(\xi)]_{{\bzero,\bzero}}$ describes the ``local density of states'' in momentum at wavevector $\xi \in \mathbb{R}^d$. This suggests the natural trace in the plane wave basis as the integral over the momentum basis, which is summarized in Theorem 3.2 from \cite{planewave2025}, which we restate:

\begin{theorem}\label{th:2.3}
    For $\epsilon$ irrational, $f \in \Lambda_{\zeta,\delta}$, and {$\hat{V}_{\bG}$} decaying exponentially in $|G_1| + |G_2|$, we have
    \begin{equation}
    \label{eq:trms}
        \TrLim f(H_\epsilon) = \frac{1}{(2\pi)^d}\int_{\mathbb{R}^d} [f(\widehat{H}_\epsilon(\xi)]_{{\bzero},{\bzero}}\,d\xi.
    \end{equation}
\end{theorem}

Finally, this can be turned into an effective algorithm by realizing that the degrees of freedom of $\widehat{H}_\epsilon(\xi)$'s contribution to $[f(\widehat{H}_\epsilon(\xi))]_{ {\bzero}, {\bzero}}$ decay in two separate parameters: lattice distance to ${\bzero}$, and wavenumber value $|\xi + G_1 + (1+\epsilon)G_2|$. To this end, we define a truncation of the degrees of freedom space
\begin{equation}
    \label{dof_space_pw}
    \WL \coloneq \Set[\Big]{ \bG \in \mathbb L^* \times \mathbb L^* \given |G_1 + (1+\epsilon)G_2| < W, \; |G_1 - (1+\epsilon)G_2| < L}.
\end{equation}
 We define the natural injection $J : \ell^2(\WL) \rightarrow \ell^2(\mathbb L^* \times \mathbb L^*)$, and $\widehat{H}_\epsilon^{\WL}(\xi) = J^* \widehat{H}_\epsilon(\xi)J$. We let $\mathcal{K}_{h}^W$ define a $h$-width discretization of $[-W,W]^d \subset \mathbb{R}^d$. Then we define an approximation of $\TrLim \; f(H_\epsilon)$ by 
 \begin{equation}
 \label{discrete_pw_dos}
    \TrLim \; f(H_\epsilon) \approx \WLDOS(f;\epsilon) = \frac{h^d}{(2\pi)^d}\sum_{\xi \in \mathcal{K}_{h}^W} [f(\widehat{H}_\epsilon^{\WL}(\xi))]_{ {\bzero},{\bzero}}.
 \end{equation}
 \begin{theorem}
 \label{thm:error:ms}
 {For $\epsilon$ irrational, $f \in \Lambda_{\zeta,\delta}$, and $\hat{V}_{\bG}$ decaying exponentially in $|G_1| + |G_2|$}. There exists some $C,c>0$ independent of $W,L,h,\zeta,$ and $\delta$ such that
     \begin{equation}
     \label{error:ms}
         |\TrLim \; f(H_\epsilon) - \WLDOS(f;\epsilon)| \leq C (\delta^{-2} e^{-c \delta L} + \delta^{-2} e^{-c \zeta W} + e^{-c \delta/h }).
     \end{equation}

 \end{theorem}
 
 Observe that {the above bound is independent of $\epsilon$. It can be made lower than $\eta > 0$ with the following scaling:
 \begin{equation}
  \label{ms:scaling}
      L \sim \delta^{-1}\big(\log(\delta^{-1})+\log(\eta^{-1})\big), \qquad W \sim \log(\delta^{-1})+\log(\eta^{-1}), \qquad h \sim \delta/\log(\eta^{-1}).
  \end{equation}
  }
Large $W$ corresponds to large momenta, which contribute weakly to lower energies due to the high kinetic energy $\frac{1}{2}|\xi|^2$. Meanwhile the decay in the degrees of freedom through $L$ are controlled only by Combes--Thomas estimates as it controls inclusion of momenta with similar kinetic energy.

\subsection{Semiclassical expansion}
\label{sec:semiclassical_expansion}

\subsubsection{Density of states of Schr\"odinger operators in the semiclassical limit}

Let us start with a gentle introduction to pseudodifferential calculus on $\R^d$ with scalar symbols, see e.g.,~\cite{dimassi1999spectral, zworski2012semiclassical}. Consider a classical Hamiltonian system on the phase space $\R^d_x \times \R^d_\xi$, where $x$ is the position variable and $\xi$ the momentum variable. Weyl calculus is a convenient way to quantize this system. It allows one to transform classical observables $a \in C^\infty(\R^d_x \times \R^d_\xi;\C)$ into quantum observables ${\rm Op}_\epsilon(a)$, that are linear operators on the position-representation quantum state space $L^2(\R^d_x;\C)$. Here $\epsilon > 0$ is a parameter, which is taken equal to the reduced Planck constant $\hbar$ in the standard setting of quantum mechanics. Formally, ${\rm Op}_\epsilon(a)$ is defined by the formula
\begin{equation}\label{eq:Weyl_scalar}
[{\rm Op}_\epsilon(a) \varphi] (x) = \frac{1}{(2\pi\epsilon)^d} \int_{\R^d_x \times \R^d_\xi} a\left( \frac{x+x'}2,\xi \right) \varphi(x')  \; e^{i \frac{\xi \cdot (x-x')}\epsilon} \, dx' \, d\xi.
\end{equation}
Clearly, ${\rm Op}_\epsilon$ is a linear map, and it is easy to check that, still formally, it maps real-valued functions into symmetric operators. In addition, it maps separable functions $a(x,\xi) = f(\xi) + g(x)$ into operators of the form ${\rm Op}_\epsilon (a)= f(-i\epsilon \nabla_x)+g(x)$. In particular if $h_{\rm cl}(x,\xi) \coloneq \frac{|\xi|^2}{2m}+V(x)$ is a classical Hamiltonian describing a point-like particle with mass $m$ subjected to an external potential $V$, then for $\epsilon=\hbar$,
$$
{\rm Op}_\epsilon (h_{\rm cl}) = - \frac{\hbar^2}{2m} \Delta_x + V(x)
$$
is the Schr\"odinger Hamiltonian describing a quantum particle with mass $m$ subjected to the external potential $V$.

\medskip

The integral representation \cref{eq:Weyl_scalar} is well-defined if, e.g., $a \in \cS(\R_x^d \times \R^d_\xi;\C)$ and $\phi \in \cS(\R^d_x;\C)$. It can be shown that if $a \in \cS(\R_x^d \times \R^d_\xi;\C)$, then ${\rm Op}_\epsilon(a)$ is a trace-class operator on $L^2(\R^d_x)$ and that
$$
\Tr\left( {\rm Op}_\epsilon(a) \right) = \frac{1}{(2\pi\epsilon)^d} \int_{\R^d_x \times \R^d_\xi} a(x,\xi) \, dx \, d\xi.
$$
Consider now a semiclassical symbol in $\cS(\R_x^d \times \R^d_\xi;\C)$, that is a smooth function $a_\bullet : (0,\epsilon_0] \ni \epsilon \mapsto a_\epsilon \in \cS(\R^d_x \times \R^d_\xi;\C)$ admitting an asymptotic expansion at any order at $0$, i.e., such that for all $n \in \N$,
$$
a_\epsilon = \sum_{j=0}^n \epsilon^j a_{(j)} + {\mathcal O}_{\cS(\R^d_x \times \R^d_\xi)}\left(\epsilon^{n+1}\right) \quad \text{with} \quad a_{(j)} \in \cS(\R_x^d \times \R^d_\xi;\C).
$$
Then $\Tr\left( {\rm Op}_\epsilon(a_\epsilon) \right)$ also has an asymptotic expansion at $0$ and
$$
\Tr\left( {\rm Op}_\epsilon(a_\epsilon) \right) =  \epsilon^{-d} \left( \sum_{j=0}^n \alpha_j  \epsilon^j  + O\left(\epsilon^{n+1}\right) \right) \quad \text{with} \quad \alpha_j \coloneq \frac{1}{(2\pi)^d} \int_{\R_x^d \times \R^d_\xi} a_{(j)}(x,\xi) \, dx \, d\xi.
$$

Assume now that $h_\bullet$ is a real-valued semiclassical symbol  in $\cS(\R_x^d \times \R^d_\xi;\R)$ and $f \in C^\infty_{\rm c}(\R;\C)$ a complex-valued compactly supported function on $\R$.
Then ${\rm Op}_\epsilon(h_\epsilon)$ is a well-defined trace-class self-adjoint operator on $L^2(\R^d_x;\C)$, and $f({\rm Op}_\epsilon(h_\epsilon))$ is a trace-class operator on $L^2(\R^d_x;\C)$.  This readily follows from the spectral theory for compact self-adjoint operators. However, $\Tr\left(f({\rm Op}_\epsilon(h_\epsilon))\right)$ is not equal to
$$
\frac{1}{(2\pi\epsilon)^d} \int_{\R^d_x \times \R^d_\xi} f(h_\epsilon(x,\xi)) \, dx \, d\xi
$$
in general, because for generic semiclassical symbols $a_\bullet$ and $b_\bullet$ in $\cS(\R_x^d \times \R^d_\xi;\C)$, the operator ${\rm Op}_\epsilon(a_\epsilon b_\epsilon)$ is not equal to the product of operators ${\rm Op}_\epsilon(a_\epsilon) {\rm Op}_\epsilon(b_\epsilon)$. Instead, it holds
$$
{{\rm Op}_\epsilon(a_\epsilon){\rm Op}_\epsilon(b_\epsilon)={\rm Op}_\epsilon(c_\epsilon)},
$$
where $c_\bullet$ is the semiclassical symbol in $\cS(\R_x^d \times \R^d_\xi;\C)$ defined as the Moyal product of the semiclassical symbols $a_\bullet$ and $b_\bullet$:
\begin{align*}
c_\epsilon(x,\xi) \coloneq{}& (a_\epsilon \#b_\epsilon)_\epsilon(x,\xi)\\
\coloneq{}& \frac{1}{(\pi\epsilon)^{2d}} \int_{(\R^{d})^4} {e^{-\frac{2i}\epsilon (\xi_1 \cdot x_2-\xi_2 \cdot x_1)}}a_\epsilon(x+x_1,\xi+\xi_1)b_\epsilon(x+x_2,\xi+\xi_2) \, dx_1\, d\xi_1 \, dx_2 \, d\xi_2 \\
={}& a_{(0)}(x,\xi) b_{(0)}(x,\xi) + \epsilon \left( a_{(1)}b_{(0)}+a_{(0)}b_{(1)} - \frac i2 \left\{a_{(0)},b_{(0)}\right\}\right) (x,\xi)  \\
& + \epsilon^2 \left( a_{(2)} b_{(0)} + a_{(1)} b_{(1)} + a_{(0)} b_{(2)} - \frac i 2 \{a_{(1)},b_{(0)}\}  - \frac i 2 \{a_{(0)},b_{(1)}\}  - \frac 18 \{a_{(0)},b_{(0)}\}_2 \right)(x,\xi)\\
&+ \cdots,
\end{align*}
where $\{ \bullet,\bullet \}$ denotes the Poisson bracket
$$
\{ f,g \} = \nabla_x f \cdot \nabla_\xi g - \nabla_\xi f \cdot \nabla_x g,
$$
and $\{ \bullet,\bullet \}_2$ the second-order Poisson bracket
$$
\{ f,g \}_2 = D^2_{xx} f : D^2_{\xi\xi} g + D^2_{\xi\xi} f : D^2_{xx} g- 2 D^2_{x\xi} f : D^2_{x\xi} g.
$$
To expand $\Tr\left( f({\rm Op}_\epsilon(h_\epsilon)) \right)$ in powers of $\epsilon$, two key ingredients are needed. First, the Helffer--Sj\"ostrand formula~\cite{Helffer1989EquationDS} allows one to rewrite the operator $f({\rm Op}_\epsilon(h_\epsilon))$ as a weighted integral over $\C$ of the resolvent $(z-{\rm Op}_\epsilon(h_\epsilon))^{-1}$: 
\begin{equation}\label{eq:Helffer-Sjostrand}
f({\rm Op}_\epsilon(h_\epsilon))=-\frac{1}{\pi}\int_{\mathbb{C}}\overline{\partial}\widetilde{f}(z)(z-{\rm Op}_\epsilon(h_\epsilon))^{-1} \, dL(z),
\end{equation}
where $z=x+iy$, $\bar\partial \coloneq \frac{1}{2}(\partial_x + i\partial_y)$, and $\widetilde f \in C^\infty_{\rm c}(\mathbb{C};\C)$ is any almost-analytic extension of $f$ satisfying
(i)~$\Supp(\widetilde{f})$ is a complex neighborhood of $\Supp(f)$,
(ii)~$\widetilde{f}(z)=f(z)$ for any $z\in \mathbb{R}$,
and (iii)~\mbox{$|\overline{\partial}\widetilde{f}(z)|=\mathcal{O}(|\Im z|^\infty)$},
i.e., for any $n\in \mathbb{N}$, \mbox{$|\overline{\partial}\widetilde{f}(z)|=\mathcal{O}(|\Im z|^n)$ when $\Im z \to 0$}.
Second, an asymptotic expansion of the resolvent $\left(z-{\rm Op}_\epsilon(h_\epsilon)\right)^{-1}$ for Weyl's quantization of semiclassical symbols in $\cS(\R^d_x \times \R^d_\xi;\R)$ can be worked out using Moyal calculus:
\begin{equation}\label{eq:expansion_resolvent}
\left(z-{\rm Op}_\epsilon(h_\epsilon)\right)^{-1} = \sum_{j=0}^{n} \epsilon^j \; {\rm Op}_\epsilon\left( r_j(z) \right) + {\mathcal O}_{\mathcal L(L^2(\R^d_x))}(\epsilon^{n+1}),
\end{equation}
with, using the fact that for scalar invertible symbols $\{a,a^{-1}\}=0$,
\begin{align*}
 r_0(z;x,\xi) \coloneq{}&(z-h_0(x,\xi))^{-1}, \\
 r_1(z;x,\xi) \coloneq{}& (z-h_0(x,\xi))^{-2} h_1(x,\xi),   \\
 r_2(z;x,\xi) \coloneq{}& - \frac 14   (z-h_0(x,\xi))^{-4} \left( \nabla_xh_0^T (D^2_{\xi\xi}h_0) \nabla_x h_0 + \nabla_\xi h_0^T (D^2_{xx}h_0) \nabla_\xi h_0 \right)(x,\xi) \\
 & \quad - \frac 14 (z-h_0(x,\xi))^{-3} \{h_0,h_0\}_2(x,\xi) + (z-h_0(x,\xi))^{-3} h_1(x,\xi)^2 \\
 & \quad + (z-h_0(x,\xi))^{-2} h_2(x,\xi).
\end{align*}

\begin{remark}
In this paper, we only consider the first three terms (orders $0$, $1$, and $2$) of the semiclassical expansions. Higher-order terms can be obtained systematically by symbolic calculus, but the number of terms in the $k$-th order term growths polynomially in $k$, which limits the practical use of the method to not too large values of $k$.
\end{remark}

Combining \cref{eq:Helffer-Sjostrand} and \cref{eq:expansion_resolvent} yields
\begin{equation} \label{eq:trfh}
\Tr \left(f({\rm Op}_\epsilon(h_\epsilon)\right)  =  \epsilon^{-d} \left( \sum_{j=0}^n f_j  \epsilon^j  + {\mathcal O}\left(\epsilon^{n+1}\right) \right)
\end{equation}
with
\begin{equation}\label{eq:f_j}
f_j \coloneq - \frac{1}{\pi (2\pi)^d} \;  \int_{\mathbb{C}} \overline{\partial}\widetilde{f}(z)  \int_{\R^d_x \times \R^d_\xi} r_j(z;x,\xi) \, dx\, d\xi \, dL(z).
\end{equation}
Using the relations
$$
f(y) = -\frac{1}{\pi} \int_\C \overline{\partial} \widetilde f(z) (z-y)^{-1} \, dL(z), \quad \frac{f^{(n)}(y)}{n!} =  -\frac{1}{\pi} \int_\C \overline{\partial} \widetilde f(z) (z-y)^{-n-1} \, dL(z),
$$
we obtain for the first three terms of the expansion
\begin{align}
    f_0 \coloneq{}& \frac{1}{(2\pi)^d} \int_{\R^d_x \times \R^d_\xi} f(h_0(x,\xi)) \, dx \, d\xi, \label{eq:zeroth_order} \\
    f_1 \coloneq{}& \frac{1}{(2\pi)^d} \int_{\R^d_x \times \R^d_\xi} f'(h_0(x,\xi)) h_1(x,\xi)\, dx \, d\xi, \label{eq:first_order}  \\
    f_2 \coloneq{}& - \frac 1{24}   \frac{1}{(2\pi)^d} \int_{\R^d_x \times \R^d_\xi} f^{(3)}(h_0(x,\xi))  \left( \nabla_xh_0^T (D^2_{\xi\xi}h_0) \nabla_x h_0 + \nabla_\xi h_0^T (D^2_{xx}h_0) \nabla_\xi h_0 \right)(x,\xi) \, dx \, d\xi  \nonumber \\
 & \quad +\frac 12 \frac{1}{(2\pi)^d} \int_{\R^d_x \times \R^d_\xi} f''(h_0(x,\xi))  \left(h_1(x,\xi)^2 - \frac 14 \{h_0,h_0\}_2(x,\xi)\right) \, dx \, d\xi \nonumber \\
 &  \quad + \frac{1}{(2\pi)^d} \int_{\R^d_x \times \R^d_\xi} f'(h_0(x,\xi)) h_2(x,\xi)\, dx \, d\xi. \label{eq:second_order}
\end{align}

Note that the above arguments do not directly apply to the usual classical Hamiltonian $h_{\rm cl}(x,\xi) = \frac{|\xi|^2}{2m}+V(x)$ as this function is not in ${\cal S}(\R^d_x \times \R^d_\xi;\R)$, even for $V \in {\cal S}(\R^d_x;\R)$ since the kinetic energy term $\xi \mapsto \frac{|\xi|^2}{2m}$ is an unbounded function. To deal with this technical difficulty, it is necessary to  work with classes of symbols larger than $\cS(\R^d_x \times \R^d_\xi;\C)$. We will not enter these technicalities here and refer the interested reader to the literature, e.g.,~\cite{zworski2012semiclassical}. Let us only mention that if $V \in {\cal S}(\R^d_x \times \R^d_\xi;\R)$, then $h_{\rm cl}$ belongs to a suitable class of symbols allowing one to given a meaning to ${\rm Op}_{\epsilon}(h_{\rm cl})$, and that ${\rm Op}_{\epsilon}(h_{\rm cl})$ is indeed equal to $-\frac{\epsilon^2}{2m}\Delta + V(x)$ as announced earlier. The latter is an unbounded essentially self-adjoint operator on $L^2(\R^d_x;\C)$ with essential spectrum the half-line $[0,+\infty)$. As a consequence, for any $f \in C^\infty_{\rm c}(\R;\C)$ with support in $(-\infty,0)$, the operator $f({\rm Op}_\epsilon(h_{\rm cl}))$ is finite-rank, hence trace-class, and it can be shown using the same techniques as above that
$$
\Tr\left( f({\rm Op}_\epsilon(h_{\rm cl})) \right) = \epsilon^{-d} \left( \sum_{j=0}^n f_j^{\rm cl}  \epsilon^j  + {\mathcal O}\left(\epsilon^{n+1}\right) \right),
$$
with the $f_j^{\rm cl}$'s given by formulae~\cref{eq:zeroth_order}-\cref{eq:second_order} with $h_0=h_{\rm cl}$ and $h_j=0$ for $j \ge 1$. As
\begin{align*}
& \nabla_xh_{\rm cl}(x,\xi)=\nabla V(x), \quad \nabla_\xi h_{\rm cl}(x,\xi)=\xi, \quad D^2_{xx}h_{\rm cl}(x,\xi) = D^2V(x), \quad D^2_{\xi\xi}h_{\rm cl}(x,\xi) = I_2, \\
& \{h_{\rm cl},h_{\rm cl}\}_2(x,\xi)=2\Delta V(x),
\end{align*}
and, by integration by parts,
\begin{align*}
&  \int_{\R^d_\xi}  f^{(3)}(h_{\rm cl}(x,\xi)) \xi_i \xi_j \, d\xi = - \delta_{ij} \int_{\R^d_\xi}  f^{(2)}(h_{\rm cl}(x,\xi))  \, d\xi, \\
&  \int_{\R^d_x}  f^{(3)}(h_{\rm cl}(x,\xi)) \left( \frac{\partial V}{\partial x_i} \right)^2 \, dx = - \int_{\R^d_x}  f^{(2)}(h_{\rm cl}(x,\xi)) \, \frac{\partial^2 V}{\partial x_i^2} \, dx,
\end{align*}
we finally get
\begin{align*}
f_0^{\rm cl} ={}& \frac{1}{(2\pi)^d} \int_{\R^d_x \times \R^d_\xi} f(h(x,\xi)) \, dx \, d\xi, \\
f_1^{\rm cl} ={}& 0, \\
f_2^{\rm cl} ={}& - \frac 1{24}  \frac{1}{(2\pi)^d} \int_{\R^d_x \times \R^d_\xi} f^{(2)}(h(x,\xi)) \, \Delta V(x) \,  \, dx \, d\xi.
\end{align*}
It can be shown that the odd terms of the expansion vanish and the even term of order $2j$ is a sum of $2j$ terms of the form $f^{(k)}(h(x,\xi)) g_{j,k}(x)$, with $j+1 \le k \le 3j$ and $g_{j,k}(x)$ is polynomial in the derivatives of $V$ at point $x$.
This expansion was obtained by Helffer and Robert in~\cite{HelfferRobert_83}, see also~\cite{ColinDeVerdiere_12}.

\subsubsection{Semiclassical analysis of the two-scale Hamiltonian \texorpdfstring{$H_\epsilon$}{Hepsilon}}

In this section, we derive the semiclassical approximations of the DoS, postponing longer computations {of \cref{eq:L0}-\cref{eq:L2}} for \Cref{sec:Appendix}, and formulating the result at the end of the section in \Cref{th:2.4}.
Let us first recall the definition of the Bloch transform with respect to the real-space lattice $\mathbb L=\Z^d$. We denote by $\Omega \coloneq \left(-\frac 12,\frac 12\right]^d$ the unit cell of $\mathbb L$, $\Omega^*=\left(-\pi,\pi\right]^d$ the first Brillouin zone (a specific unit cell of the dual lattice $\mathbb L^*=2\pi\Z^d$),
\[
  L^2_{\rm per}(\Omega)  \coloneq \Set*{ u \in L^2_{\rm loc}(\R^d_x;\C) \given u \text{ is  $\Z^d$-periodic} },
\]
and
\[
  \cH \coloneq
  \Set*{
    v \in L^2_{\rm loc}(\R^d_\xi;L^2_{\rm per}(\Omega))
    \given
    v_{k+G}(x) = e^{iG \cdot x}v_k(x) \text{ for all } G \in \mathbb L^* \text{ and a.e. } (k,x) \in \R^d_\xi \times \R^d_x
  }.
\]
The space $\cH$ is endowed with the inner product
$$
\langle u_\bullet ,v_\bullet \rangle_\cH \coloneq \fint_{\Omega^*} \langle u_k, v_k \rangle_{L^2_{\rm per}}  \, dk
$$
where
\begin{align*}
    \langle u,v \rangle_{L^2_{\rm per}} \coloneq \int_{\Omega} u^*(x)v(x) \, dx .
\end{align*}

The Bloch transform is the unitary operator $\cU:L^2(\R_x;\C) \to \cH$ such that
$$
\forall \phi \in C^\infty_{\rm c}(\R^d_x;\C), \quad (\cU\phi)_k(x) = \sum_{R \in \mathbb L} \phi(x+R) e^{-ik \cdot (x+R)}.
$$
For each $(k,X) \in \R^d_\xi \times \R^d_x$, we denote by $h(k,X)$ the self-adjoint operator on $L^2_{\rm per}(\Omega)$ with domain $H^2_{\rm per}(\Omega)$ defined by
$$
\forall \phi \in H^2_{\rm per}(\Omega), \quad (h(k,X)\phi)(x) \coloneq  \frac 12 \left[\left( -i \nabla+k \right)^2 \phi\right](x) + V(x,x+X)\phi(x).
$$
The variable $X$ is local disregistry which represents $\epsilon x$. For any fixed $X$, the operator $h(k,X)$ can be regarded as a Bloch decomposition of the commensurate approximation at this disregistry:
\begin{align*}
 \mathfrak{h}(X) \coloneq -\frac{1}{2}\Delta+V(x,x+X).
\end{align*}
From this point of view, $H_\epsilon$ for $\epsilon \rightarrow 0$ is formally a uniform union of $\mathfrak{h}(X)$ following the intuition of Birkhoff's ergodic theorem.
The semiclassical analysis of the two-scale Hamiltonian \cref{eq:Hepsilon} is based on the observation that
\begin{equation} \label{eq:WeylHepsilon}
H_\epsilon = \cU^{-1} {\rm Op}_\epsilon(h) \cU,
\end{equation}
where ${\rm Op}_\epsilon(h)$ is the Weyl quantization of the operator-valued symbol $h$. If $a(k,X)$ is a symbol with values in the space of operators on $L^2_{\rm per}(\Omega)$, ${\rm Op}_\epsilon(a)$ is the operator on $\mathcal H$ formally defined by
$$
[{\rm Op}_\epsilon(a) \phi]_k (x) \coloneq \frac{1}{(2\pi\epsilon)^d} \int_{\R^d_\xi \times \R^d_x} \left[a\left( \frac{k+k'}2,X \right) \phi_{k'} \right](x) \; e^{-i \frac{(k-k')\cdot X}\epsilon} \, dk' \, dX.
$$
Proceeding as in \cite{CancesLong}, we obtain that for all $f \in C^\infty_{\rm c}(\R;\R)$ and all $n \in \N$
$$
\begin{aligned}
\TrLim[f(H_\epsilon)]=\sum_{j=0}^n \frac{\epsilon^j}{(2\pi)^d}  \int_{\Omega}\int_{\Omega^*}\Tr_{L^2_{\rm per}}[f_j(k,X)] \, dk \, dX+\mathcal{O}(\epsilon^{n+1}),
\end{aligned}
$$
with
\begin{align}
    f_{0}(k,X)& \coloneq f(h(k,X)), \label{eq:f0}\\
         f_{1}(k,X)& \coloneq -\frac{1}{\pi}\int_\C \overline{\partial}\widetilde{f}(z)  \left( \frac i2 (z-h)^{-1} \{(z-h),(z-h)^{-1}\} \right)(k,X) \, dL(z),  \label{eq:f1} \\
         f_{2}(k,X)& \coloneq -\frac{1}{\pi}\int_\C \overline{\partial}\widetilde{f}(z)  \Big(-\frac 14 (z-h)^{-1}\{(z-h),(z-h)^{-1}\{(z-h),(z-h)^{-1}\}\} \notag\\
     &\hspace{31.5mm}+\frac{1}{8}(z-h)^{-1}\{(z-h),(z-h)^{-1}\}_2\Big)(k,X) \, dL(z),  \label{eq:f2}\\
         \cdots\notag
\end{align}
 where $\widetilde{f}: \C \to \C$ is any almost analytic extension of $f$, and $\{\bullet,\bullet\}$ and $\{\bullet,\bullet\}_2$ are the Poisson and second-order Poisson brackets respectively defined in this setting by
 \begin{align*}
   \{a,b\} \coloneq{}& \nabla_X a \cdot \nabla_k b-\nabla_k a \cdot \nabla_X b, \\
   \{a,b\}_{2} \coloneq{}& D^2_{kk}a : D^2_{XX}b + D^2_{XX}a : D^2_{kk}b-2 D^2_{kX} a:  D^2_{kX}b.
\end{align*}

\medskip

In contrast with the scalar scale dealt with in the previous section, the expressions of $f_1$, $f_2$, \ldots{} are less explicit, essentially because operator-valued symbols do not commute. They can however be rewritten as sum-over-states formulae. To do this, let us first diagonalize each compact-resolvent self-adjoint operator $h(k,X)$ in an orthonormal basis, i.e.,
\begin{align}\label{eq:h-decomp}
    h(k,X) = \sum_{n=1}^{+\infty} \lambda_n(k,X) |u_n(k,X)\rangle \langle u_n(k,X)|, \qquad  \langle u_m(k,X)|u_n(k,X) \rangle = \delta_{mn}.
\end{align}
Introducing the matrix elements
\begin{align*}
\cK_{mn}(k,X) \coloneq{}& \langle u_m(k,X)|\nabla_kh(k,X)|u_n(k,X)\rangle =  \langle u_m(k,X)|- i \nabla +k |u_n(k,X)\rangle \\
={}&  \langle u_m(k,X)|- i \nabla |u_n(k,X)\rangle + k \delta_{mn}, \\
\cX_{mn}(k,X) \coloneq{}& \langle u_m(k,X)|\nabla_X h(k,X)|u_n(k,X)\rangle = \langle u_m(k,X)|\nabla_X V(\cdot,X) |u_n(k,X)\rangle, \\
\cX_{mn}^{(2)}(k,X) \coloneq{}& \langle u_m(k,X)|D^2_{XX} h(k,X)|u_n(k,X)\rangle = \langle u_m(k,X)|D^2_{XX}V(\cdot,X) |u_n(k,X)\rangle,
\end{align*}
(note that $\cK_{mn}^{(2)}(k,X) \coloneq \langle u_m(k,X)|D^2_{kk} h(k,X)|u_n(k,X)\rangle = \delta_{mn}$), the finite differences
\begin{align*}
f^{(2)}_2(\lambda;\lambda') \coloneq{}& - \frac {2!} \pi \int_\C \frac{\overline{\partial}\widetilde{f}(z)}{(z-\lambda)^2(z-\lambda')}  \, dL(z) = \left| \begin{array}{ll} 2 \frac{f(\lambda')-f(\lambda)- (\lambda'-\lambda)f'(\lambda)}{(\lambda'-\lambda)^2} \quad & \text{if $\lambda\neq\lambda'$} \\
  f^{(2)}(\lambda) \quad & \text{if $\lambda=\lambda'$} \end{array} \right., \\
 f^{(3)}_3(\lambda;\lambda',\lambda'') \coloneq{}& - \frac {3!} \pi \int_\C \frac{\overline{\partial}\widetilde{f}(z)}{(z-\lambda)^2(z-\lambda')(z-\lambda'')}  \, dL(z) \\
={}& \left| \begin{array}{ll} 3 \frac{f^{(2)}_2(\lambda;\lambda'')-f^{(2)}_2(\lambda;\lambda')}{\lambda''-\lambda'}  \quad & \text{if $\lambda'\neq\lambda''$} \\
 3 \frac{\partial f^{(2)}_2}{\partial\lambda'}(\lambda;\lambda') = -12
\frac{f(\lambda')-f(\lambda) - \frac 12 (f'(\lambda')+f'(\lambda))(\lambda'-\lambda)}{(\lambda'-\lambda)^3}  \quad  & \text{if $\lambda\neq\lambda'=\lambda''$} \\
f^{(3)}(\lambda) & \text{if $\lambda=\lambda'=\lambda''$} \end{array} \right., \\
f^{(4)}_4(\lambda;\lambda',\lambda'',\lambda''') \coloneq{}& - \frac {4!} \pi \int_\C \frac{\overline{\partial}\widetilde{f}(z)}{(z-\lambda)^2(z-\lambda')(z-\lambda'')(z-\lambda''')}  \, dL(z) \\
={}& \left| \begin{array}{ll} 4 \frac{f^{(3)}_3(\lambda;\lambda',\lambda''')-f^{(3)}_3(\lambda;\lambda',\lambda'')}{\lambda'''-\lambda''}  \quad & \text{if $\lambda''\neq\lambda'''$} \\
 4 \frac{\partial f^{(3)}_3}{\partial\lambda''}(\lambda;\lambda',\lambda'')   \quad & \text{if $\lambda'\neq \lambda''=\lambda'''$} \\
  6\frac{\partial^2 f^{(2)}_2}{(\partial\lambda'')^2}(\lambda;\lambda')   \quad & \text{if $\lambda'= \lambda''=\lambda'''$} \\
 f^{(4)}(\lambda) & \text{if $\lambda=\lambda'=\lambda''=\lambda'''$}
\end{array} \right.,
\end{align*}
and the linear forms $L_j: C^0_{\rm c}(\R;\C) \to \C$, $j=0,1,2$, defined by
\begin{align}
L_0(f) \coloneq{}& \frac{1}{(2\pi)^d} \int_\Omega \int_{\Omega^*} \sum_{n=1}^{+\infty} f(\lambda_n(k,X)) \, dk \, dX, \label{eq:L0}\\
L_1(f) \coloneq{}&-\frac{1}{(2\pi)^d} \int_\Omega \int_{\Omega^*} \sum_{m,n=1}^{+\infty} \frac 12 f^{(2)}_2(\lambda_m(k,X),\lambda_n(k,X))  \Im \left( \cK_{mn}(k,X) \cdot \cX_{nm}(k,X) \right) \, dk \, dX, \label{eq:L1} \\
L_2(f) \coloneq{}& \frac{1}{(2\pi)^d} \int_\Omega \int_{\Omega^*} \bigg( - \frac 14  \sum_{m} \frac{f^{(2)}(\lambda_m)}{2!} \Tr(\cX^{(2)}_{mm})\notag \\
& \quad - \frac 14  \sum_{m,n} \frac{2 f^{(3)}_3(\lambda_m;\lambda_m,\lambda_n)-f^{(3)}_3(\lambda_m;\lambda_n,\lambda_n)}{3!}  |\cX_{mn}|^2 \notag \\
& \quad - \frac 14 \sum_{m,n,p} \frac{f^{(3)}_3(\lambda_m;\lambda_n,\lambda_p)}{3!} \left( 2 \cK_{np} \cdot \cX^{(2)}_{mn} \cK_{pm}  -  \cK_{mn} \cdot \cX^{(2)}_{np} \cK_{pm} \right) \notag \\
& \quad + \frac 14   \sum_{m,n,p,q} \frac{f_4^{(4)}(\lambda_m;\lambda_n,\lambda_p,\lambda_q)}{4!} \Big[(\mathcal{X}_{mn}\cdot\mathcal{K}_{np})(\mathcal{K}_{pq}\cdot\mathcal{X}_{qm})\notag\\
&\qquad+(\mathcal{X}_{mn}\cdot \mathcal{K}_{pq})(\mathcal{K}_{np}\cdot \mathcal{X}_{qm})+(\mathcal{K}_{mn}\cdot \mathcal{X}_{pq})(\mathcal{X}_{np}\cdot\mathcal{K}_{qm})+(\mathcal{K}_{mn}\cdot\mathcal{X}_{np})(\mathcal{X}_{pq}\cdot\mathcal{K}_{qm})\notag \\
&\qquad -2\Re\Big((\mathcal{K}_{mn}\cdot \mathcal{X}_{np})(\mathcal{K}_{pq}\cdot\mathcal{X}_{qm})\Big)-2\Re \big((\mathcal{X}_{mn}\cdot \mathcal{K}_{pq})(\mathcal{X}_{np}\cdot \mathcal{K}_{qm})\big)\notag\\
&\qquad +2\Re \big((\mathcal{X}_{mn}\cdot \mathcal{K}_{qm})(\mathcal{K}_{np}\cdot \mathcal{X}_{pq}-\mathcal{X}_{np}\cdot \mathcal{K}_{pq})\big)\Big] \bigg)(k,X)dkdX,
 \label{eq:L2}
\end{align}
we finally obtain the following result:
\begin{theorem}\label{th:2.4}
Assume that $V \in C^\infty(\R^d_x \times \R^d_x;\R)$ is an $\mathbb L$-periodic function with respect to each variable. Then for {each $m\in \N$, there exist $C_{d,m}>0$ such that for all $f\in C^\infty_c(\R;\C)$ and $\epsilon > 0$ small enough, we have that}, 
\begin{equation}
\label{tr2ord-error_2}
    \left|\TrLim[f(H_\epsilon)] -\sum_{j=0}^m \epsilon^j L_j(f)\right|
    \leq C_{d,m} \epsilon^{m+1}\sum_{n\leq 2m+6d+8}\left\|\frac{d^{n}f}{ds^{n}}(s)\right\|_{L^\infty(\R)},
\end{equation}
where $L_0(f)$, $L_1(f)$, $L_2(f)$ are given by \cref{eq:L0}-\cref{eq:L2}, and $L_j(f)$, $j \ge 3$ can be computed using derivatives of $V$ up to order $j$ and derivatives of $f$ up to order $2j$.
\end{theorem}

{While a theoretical error analysis is available for expansions of arbitrary order in \Cref{th:2.4}, a numerical verification of the convergence rates remains practically challenging. This is due to the rapid increase in complexity involved in deriving computable expressions for $L_j(f)$ in terms of $\{(\lambda_n(k,X), u_n(k,X))\}$. In particular, higher-order terms in the Moyal product involve higher-order bidifferential operators (referred to here as higher-order Poisson brackets), for which obtaining computable formulas becomes more difficult at higher orders. Nonetheless, for small enough values of $\epsilon$, the second-order approximation
} 
\begin{equation} \label{tr2ord-error}
    \TrLim[f(H_\epsilon)] \approx L_0(f)+\epsilon L_1(f) + \epsilon^2 L_2(f)
\end{equation}
{provides a good balance between theoretical accuracy and computational efficiency. Accordingly, we employ this second-order expansion for the numerical experiments presented in the next section.} 

\begin{remark}
    The class $C^\infty_c(\R;\C)$ of test functions in \Cref{th:2.4} is different from the class $\Lambda_{\zeta,\delta}$ (defined in~\cref{def:Lambda-zeta-delta}) of test functions in \Cref{th:2.3}. However, if $\chi\in C^\infty_c(\R;[0,1])$ is such that
    \[
    \chi(x) =  \left\{\begin{array}{ll} 1, \quad&|x|\leq\frac{1}{3}\\[1ex]
    0,&|x|>1\end{array}\right.
    \qquad {\rm and} \qquad \sum_{R\in \Z} \chi(x-R)=1,
    \]
then for any $f\in C^\infty(\R;\C)$, \Cref{th:2.4} holds piecewise for $\chi(\cdot-R)f$ with $R\in \Z$. This implies that \Cref{th:2.4} holds at order $m$ for any $f\in C^\infty(\R;\C)$ such that
\begin{align}\label{eq:2.28}
  \sum_{R\in \Z}  \sum_{n\leq 2m+6d+8}\left\|\frac{d^n}{dx^n}f(x)\right\|_{L^\infty([-1,1)+R)}<\infty.
\end{align}
In particular, this condition is satisfied for any order $m$ by the Gaussian test functions used in the numerical simulations reported in \Cref{sec:numerics}.
\end{remark}

\section{Numerical experiments}
\label{sec:numerics}

In this section, we numerically study the density of states of \cref{eq:Hepsilon} using the momentum-space and semiclassical methods. {These two methods are, in principle, applicable in any dimension and converge in the sense of \Cref{thm:error:ms,th:2.4}. However, their asymptotic scaling and underlying mechanisms are completely different. For the momentum-space method, by replacing the eigensolver in \cref{discrete_pw_dos} with the kernel polynomial method (KPM) \cite{weisse2006kernel}, the computational scaling is $\mathcal{O}(\delta^{-1} h^{-d}L^dW^{2d})$, where $\delta$ denotes the analytic extension width of the test functions. Consequently, the computational complexity of the momentum-space method is independent of the incommensurability parameter $\epsilon$, but relies heavily on the analytical properties of the test function (see also \cref{ms:scaling}). For the semiclassical method, by treating the diagonalization of $h(k,X)$ as a fixed pre-processing step, the scaling is $\mathcal{O}(N_{\rm bands}^{2m})$, where $m$ is the expansion order and $N_{\rm bands}$ the number of bands retained at each phase-space point $(k,X) \in \Omega^*\times \Omega$. 
To better explore these two methods across different regimes (e.g., varying values of $\epsilon$ and the analyticity or smoothness of the test functions), we restrict our numerical experiments to the one-dimensional case. Since the computational cost of higher-dimensional simulations is significantly higher, this restriction allows us to obtain highly accurate results within a feasible timeframe.} 

We consider the one-dimensional toy model,
\begin{equation*}
  H_{\epsilon} = -\frac{1}{2}  \frac{d^2}{dx^2} + V(x,(1+\epsilon)x) \quad  \mbox{on } L^2(\R),
\end{equation*}
with an external potential~\( V \) that is the superposition of two Gaussian functions
\begin{equation*}
  V(x,(1+\epsilon)x) \coloneq \sum_{R\in \Z} \Big(v_1\big(x-R\big)+v_2\big((1+\epsilon)x-R\big)\Big),
\end{equation*}
where for~\( j \in \{1, 2\} \)
\begin{equation*}
  v_j(x) = -A_{j} \delta_{\sigma_{j}}(x)
  \qquad \text{with} \qquad
  \delta_{\sigma}(x) \coloneq \frac{1}{\sigma\sqrt{2\pi}}e^{-x^2/2\sigma^2}.
\end{equation*}
We have chosen Gaussian functions $v_1$ and $v_2$ with amplitudes $A_1 = 7.0$ and $A_2 = 5.0$ and variance $\sigma_1=\sigma_2 = 0.05$, as this choice provided us with an operator-valued symbol $h(k,X)$ with band gaps and energy levels that vary noticeably with respect to the \( k \)-points and \( X\)-disregistries (see \Cref{fig:lower_bands}). { This model can be considered as a moir\'e material constructed by strain. For more physical meaningful moir\'e models generated by strain, we refer to \cite{Dirac-Harper,sinner2023strain,becker2024semiclassical,one-dimensionalmoirechains} and reference therein.}

In the following, we first compare the approximations of $\TrLim(f(H_\epsilon))$ obtained by the two methods for the family of Gaussian test functions
\begin{equation}\label{eq:testfunction}
  f = \delta_\sigma(E-\bullet),
\end{equation}
{where $E$ scans the energy window of interest, and the smearing parameter $\sigma$ characterizes the width of analytic extension, while its inverse, $\sigma^{-1}$, governs the magnitude of the derivative bounds.} We will thus plot numerical approximations of the regularized density of states 
\begin{equation}
    E \mapsto \nu_{\epsilon,\sigma}(E) \coloneq \TrLim(\delta_\sigma(E-H_\epsilon)) = \left( \nu_\epsilon \star \delta_\sigma \right)(E).
\end{equation}
We then analyze the oscillations around the band gaps observed in the momentum-space results, and show that these oscillations can be described for small enough values of $\epsilon$ by an effective Hamiltonian derived from a semiclassical analysis.

\subsection{Discretization}
\label{sec:discretization}

\subsubsection{Momentum-space method}
To compute the DoS using the momentum-space method, we employ the discretization scheme defined in \cref{discrete_pw_dos}. The {KPM} is applied to avoid solving eigenvalue problems. We will consider different Gaussian smearing parameters $\sigma$ in \cref{eq:testfunction} by taking $\sigma=0.4, 0.08, 0.04$. To achieve numerical convergence in momentum-space computations, the following discretization parameters are used: $W=80, L=4000, h=0.05$ for $\sigma=0.4$, $W=80, L=5000, h=0.01$ for $\sigma=0.08$, and $W=80, L=6000, h=0.005$ for $\sigma=0.04$, respectively.

\subsubsection{Semiclassical expansion}

To compute the semiclassical expansion in \cref{tr2ord-error}, we discretize
\begin{itemize}
\item the unit cell~\( \Omega = (-\frac{1}{2}, \frac{1}{2}]  \) with a finite number of $X$-disregistries;
\item the Brillouin zone~\( \Omega^{*} = (-\pi, \pi] \) with a finite number of $k$-points;
\item each operator~\( h(k, X) \) on a planewave basis
  \begin{equation}
    \Set*{ e^{iG\bullet} \given (G + k)^{2}/2 < E_{\textrm{cut}},\;\; G \in \mathbb{L}^{*}},
  \end{equation}
  where~\( E_{\textrm{cut}} \) is an energy cutoff.
\end{itemize}
For the rest of the section, we considered \num{10 000} equally-spaced $X$-disregistries, \num{1 000} equally-spaced $k$-points and an energy cutoff of \num{10 000}. The numerical results reported below have been obtained with our implementation of the semiclassical method in the DFTK package.

\subsection{Density of states calculations}
\label{sec:DOS}

\begin{figure}[htbp]
    \centering
    \includegraphics[scale=0.5]{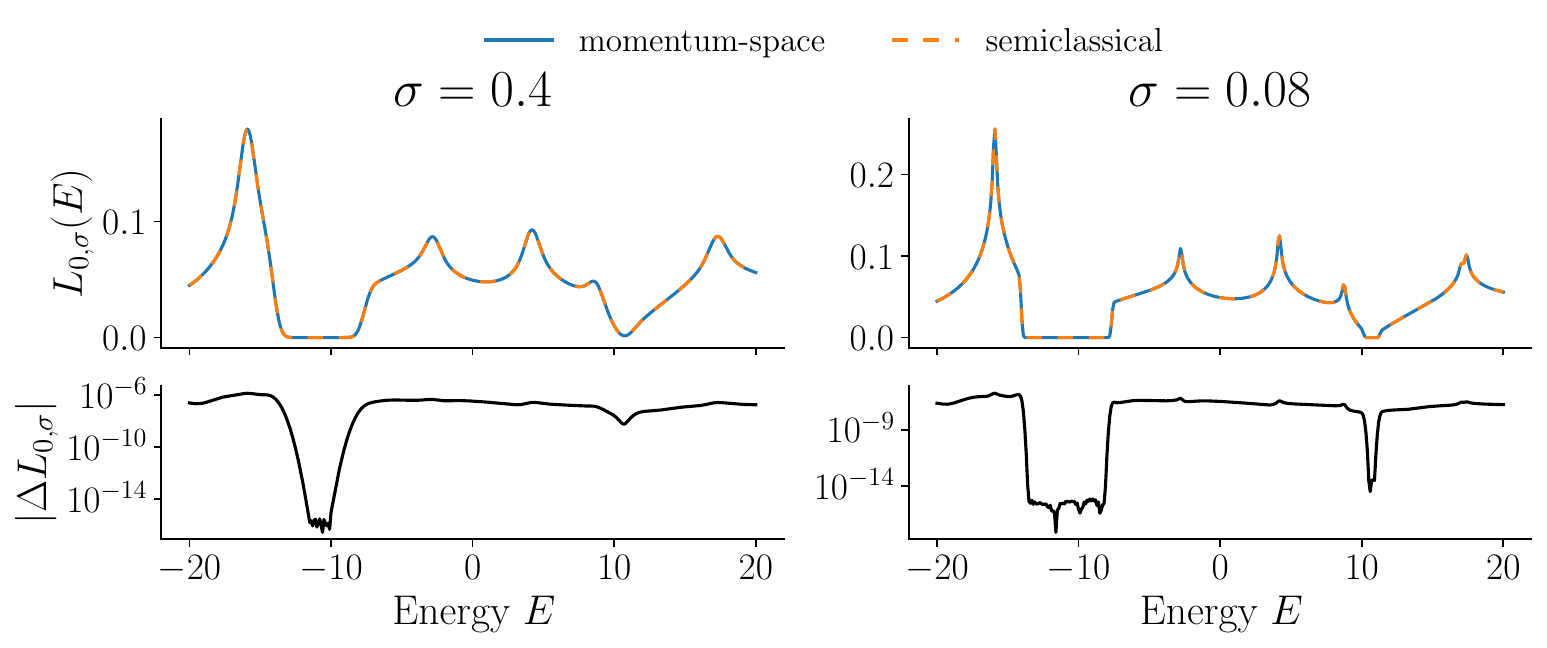}
    \caption{Comparison of two numerical approximations of the function $E \mapsto L_{0,\sigma}(E)$ over the energy range \numrange{-20}{20} for $\sigma=0.4$~(Left) and $\sigma=0.08$~(Right). The approximations obtained by the momentum-space method are in solid blue, the ones obtained by discretization of the semiclassical formulae are in dashed orange, and the absolute errors between the two approximations are in black.}\label{fig:0order}
\end{figure}

\begin{figure}[htbp]
    \centering
    \includegraphics[scale=0.5]{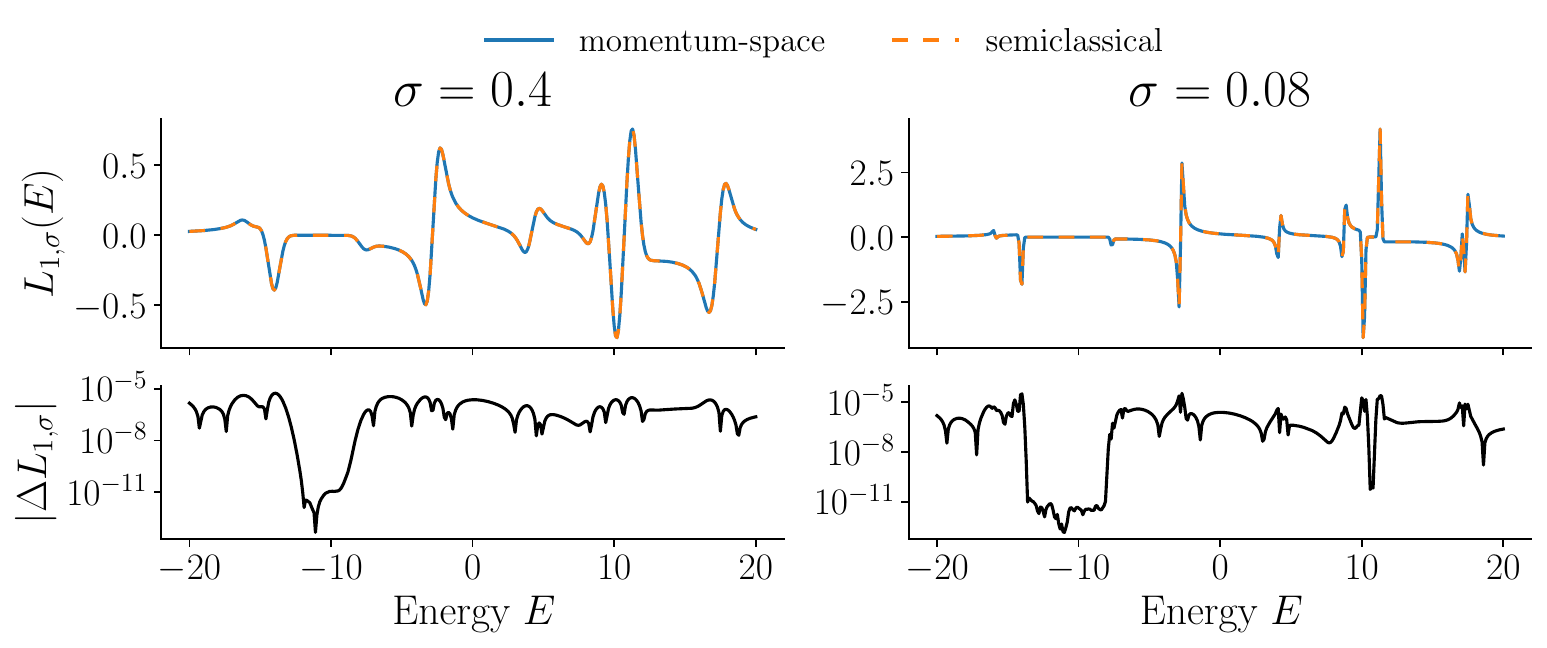}
    \caption{Comparison of two numerical approximations of the function $E \mapsto L_{1,\sigma}(E)$ over the energy range \numrange{-20}{20} for $\sigma=0.4$~(Left) and $\sigma=0.08$~(Right). The approximations obtained by the momentum-space method are in solid blue, the ones obtained by discretization of the semiclassical formulae are in dashed orange, and the absolute errors between the two approximations are in black.}\label{fig:1order}
\end{figure}

\begin{figure}[htbp]
    \centering
    \includegraphics[scale=0.5]{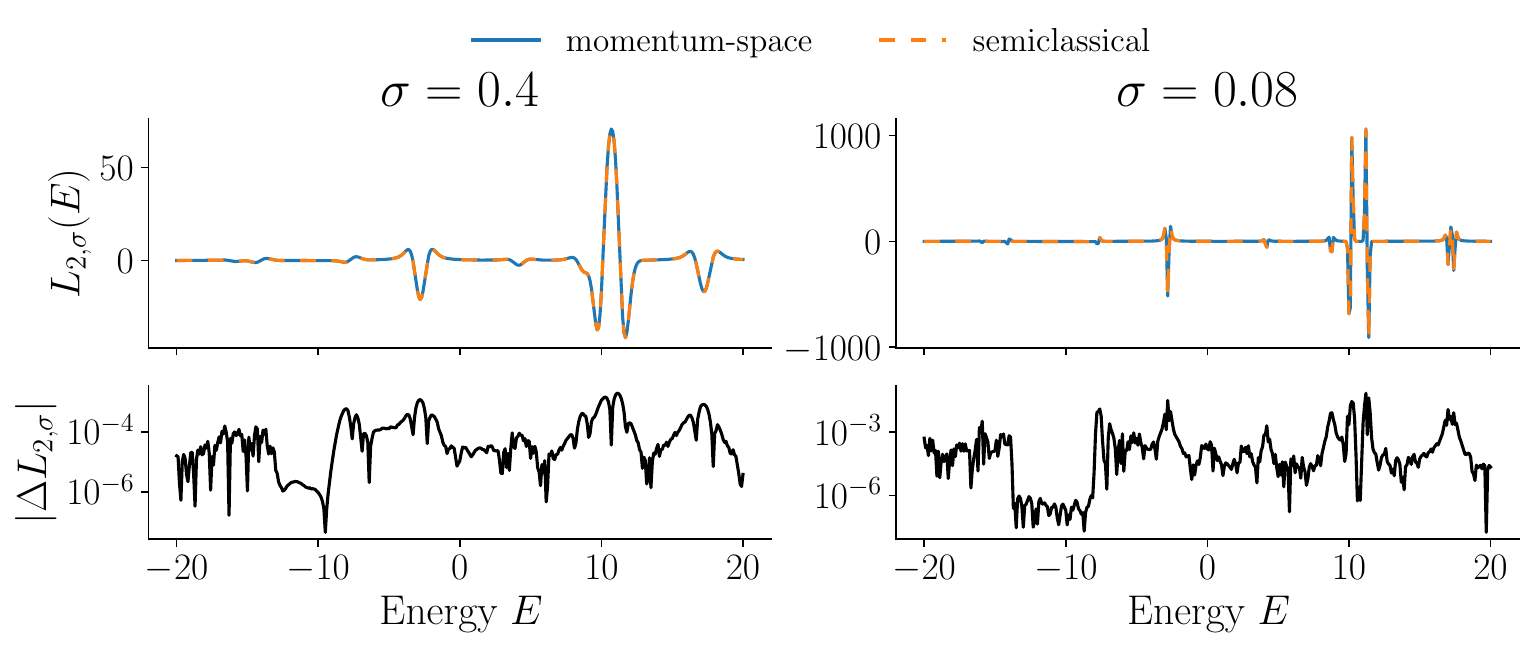}
    \caption{Comparison of two numerical approximations of the function $E \mapsto L_{2,\sigma}(E)$ over the energy range \numrange{-20}{20} for $\sigma=0.4$~(Left) and $\sigma=0.08$~(Right). The approximations obtained by the momentum-space method are in solid blue, the ones obtained by discretization of the semiclassical formulae are in dashed orange, and the absolute errors between the two approximations are in black.}\label{fig:2order}
\end{figure}

To check the correctness of the semiclassical expansion, we first compare the asymptotic expansion terms 
$$
L_{j,\sigma}(E) \coloneq L_j(\delta_\sigma(E)) = \left.\frac{\partial^j \nu_{\epsilon,\sigma}(E)}{\partial \epsilon^j}\right|_{\epsilon=0} = \left(  \left.\frac{\partial^j \nu_{\epsilon}}{\partial \epsilon^j}\right|_{\epsilon=0} \star \delta_\sigma \right)(E), \quad j=0,1,2
$$
$\sigma = 0.4$ and $\sigma = 0.08$, respectively, with  momentum-space approximations of these functions obtained from a finite‐difference approximation in $\epsilon$ of the regularized DoS $\nu_{\epsilon,\sigma}$ computed from \cref{discrete_pw_dos}. As shown in \Cref{fig:0order,fig:1order,fig:2order}, the semiclassical and momentum-space approximations are in excellent agreement over the entire energy range.

\begin{figure}[htbp]
    \centering
    \includegraphics[scale=0.5]{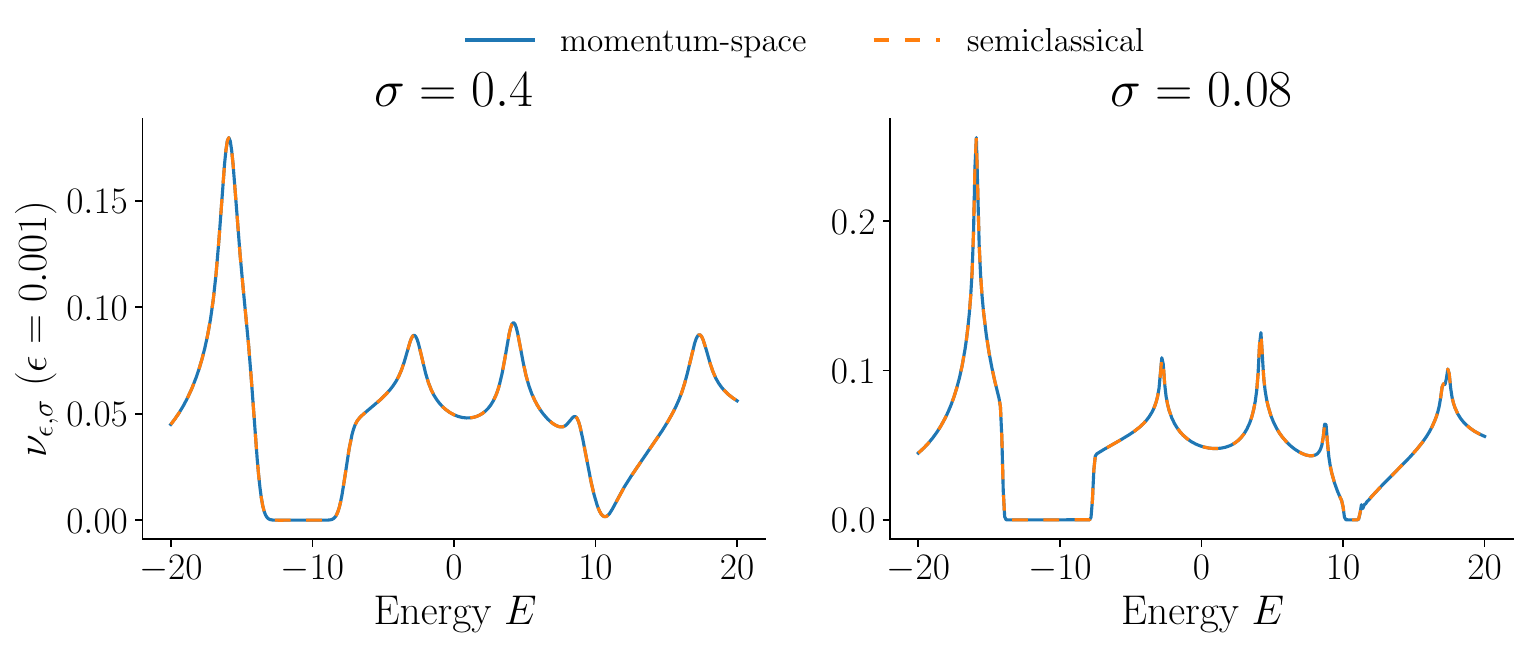}
    \caption{Comparison of the momentum space (solid blue) and second-order semiclassical (dashed orange) approximations of $\nu_{\epsilon,\sigma}$ of $\epsilon=0.001$ over the energy range \numrange{-20}{20} for $\sigma=0.4$~(Left) and ${\sigma=0.08}$~(Right).}
    \label{fig:dos:small}
\end{figure}

\begin{figure}[htbp]
    \centering
    \includegraphics[scale=0.5]{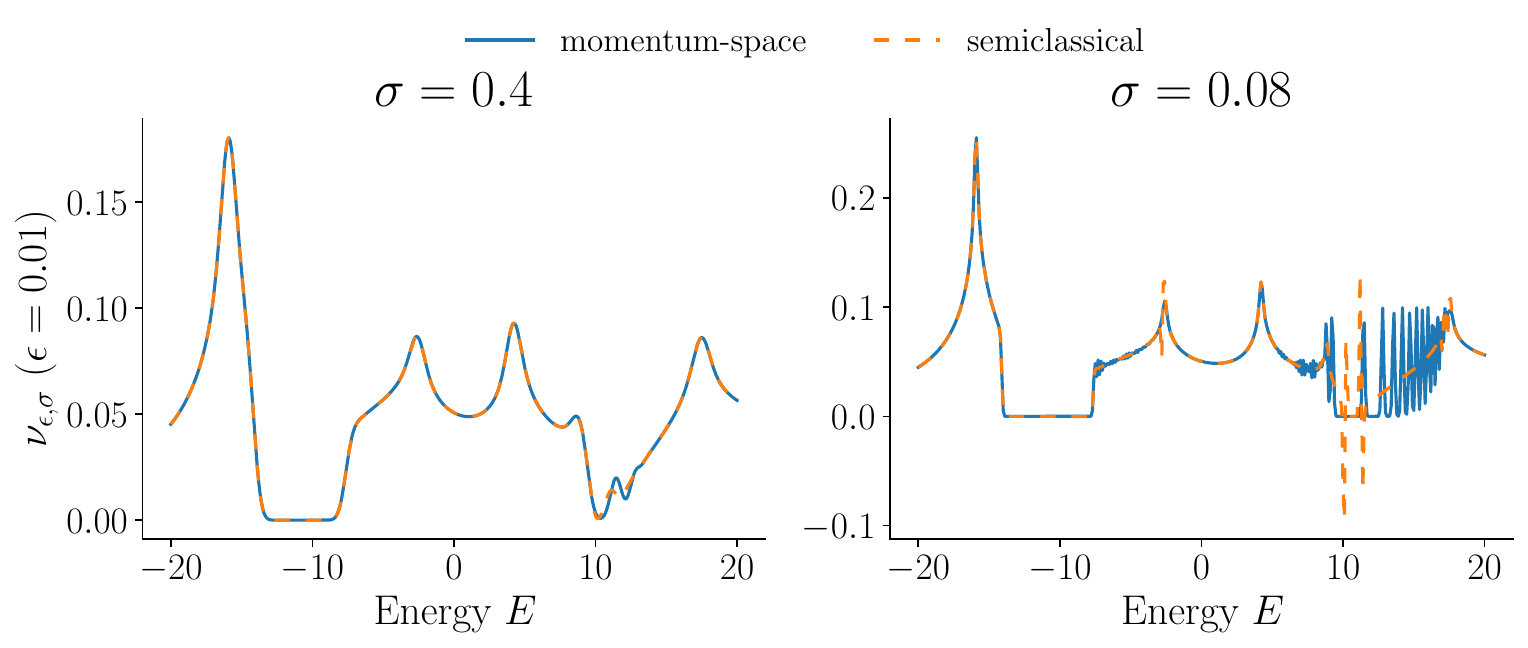}
    \caption{Comparison of the momentum space (solid blue) and second-order semiclassical (dashed orange) approximations of $\nu_{\epsilon,\sigma}$ of $\epsilon=0.01$ over the energy range \numrange{-20}{20} for $\sigma=0.4$~(Left) and ${\sigma=0.08}$~(Right).}
    \label{fig:dos:large}
\end{figure}

Next, we compare the approximations of the regularized density of states $\nu_{\epsilon,\sigma}$ obtained with \cref{discrete_pw_dos} and \cref{tr2ord-error} for two different values of $\epsilon$ ($\epsilon=0.001$ and $\epsilon=0.01$) and two different values of $\sigma$ ($\sigma=0.4$ and $\sigma=0.08$).
In \Cref{fig:dos:small}, we observe that when $\epsilon$ is small enough (here $\epsilon=0.001$), there is strong agreement across the entire energy range.
We further compare the momentum space/semiclassical approximations of $\nu_{\epsilon,\sigma}$ for a less small value of~$\epsilon$ ($\epsilon=0.01$) in \Cref{fig:dos:large}.
We see that for $\sigma = 0.4$, slight differences appear around an energy value of~10; for $\sigma = 0.08$, significant differences occur in the range $[9,18]$, along with minor differences in other energy ranges.
In addition, we note that the approximations of $\nu_{\epsilon,\sigma}$ calculated using the momentum-space method exhibit pronounced oscillations in some energy ranges.
The functions $L_{1,\sigma}$ and $L_{2,\sigma}$ displayed in 
\Cref{fig:1order,fig:2order} also have dramatic changes in those energy ranges.

\subsection{Effective model for oscillations in DoS}\label{sec:effective_moire_scale_Hamiltonians}

Although the second-order semiclassical method fails to capture the fast oscillations of $\nu_{\epsilon,\sigma}$,  the semiclassical approach provides a qualitative, and even quantitative in the limit $|\epsilon|\ll 1$, characterization of these oscillations.

\begin{figure}[htbp]
  \centering
    \subfloat[Oscillations of $\nu_{\epsilon,\sigma}$]{\label{fig:oscillations}\includegraphics[scale=0.5]{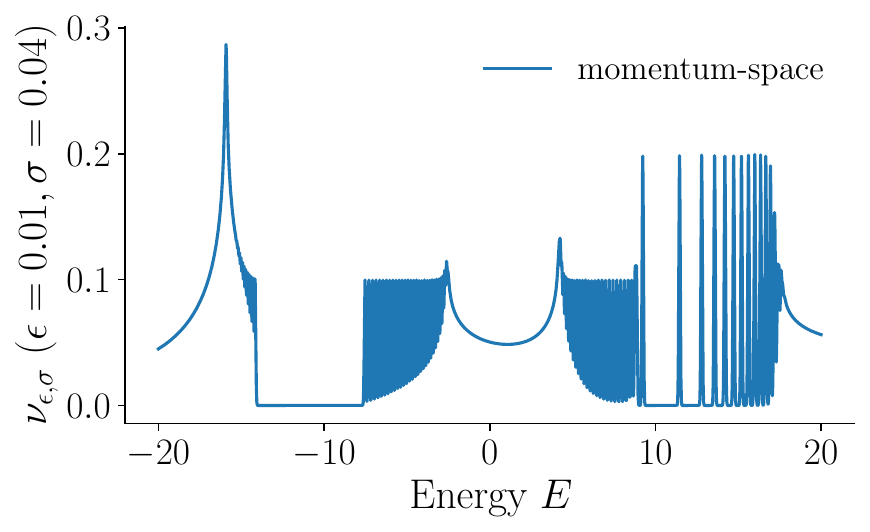}}
    \subfloat[Band structures]{\label{fig:lower_bands}\includegraphics[scale=0.5]{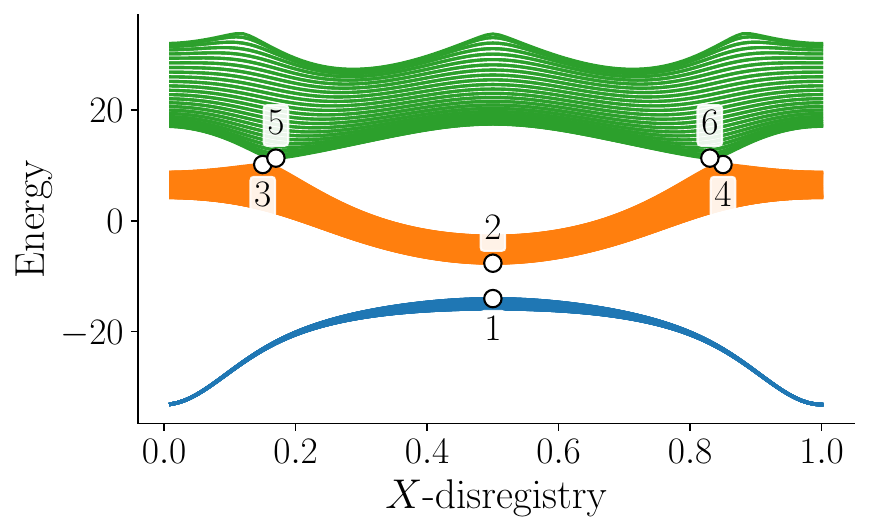}}
    \caption{(a)~Approximation of {$\nu_{\epsilon,\sigma}$ obtained with} the momentum-space method over the energy range \numrange{-20}{20} with $\epsilon=0.01$ and $\sigma=0.04$.
      (b)~First three energy bands~$E_{j}(k,X)$ of the operator $h(k,X)$.
      Each color represents a different band ($j =  1, 2,3$), and lines of the same color correspond to different $k$-points.
      The circles labeled \numrange{1}{6} mark critical points $p$.}
\end{figure}

Let $E_j(k,X)$ be the $j$-th eigenvalue of the operator $h(k,X)$. As shown in \Cref{fig:oscillations}, the fast oscillations of $\nu_{\epsilon,\sigma}$ are localized in energy ranges close to critical values of the functions $E_j(k,X)$, i.e., close to $E_j(k_0,X_0)$ where $(k_0,X_0) \in \Omega^*\times \Omega$ satisfies
$$
\nabla_k E_j(k_0,X_0) = 0 \quad \mbox{and} \quad \nabla_X E_j(k_0,X_0) = 0 \quad \mbox{for some } j.
$$
These values are the analogue of Van Hove singularities in periodic systems~\cite{VanHove}. {In physics,  Van Hove singularities have been also considered in incommensurate systems such as twisted bilayer graphene, see \cite{li2010observation}}.

For the sake of simplifying notation, we fix the band structure index $j$ and set $E(k,X) \coloneq E_j(k,X)$ in the following discussion. Let $P \subset \Omega^*\times \Omega$ be the set of the coordinates of the critical points of the band structures $(k,X)\mapsto E(k,X)$. For each $p \coloneq (k_0,X_0)\in P$, we have
\begin{equation}\label{eq:Morser}
    E(k_0 + \kappa, X_0 + Y)
    \approx E(k_0,X_0)+ \frac{1}{2}(A_p\kappa^2+ 2 B_p\kappa Y+ C_pY^2),
\end{equation}
where
\begin{align*}
  A_p \coloneq \frac{\partial^2 E}{\partial k^2}(k_0,X_0), \quad
  B_p \coloneq \frac{\partial^2 E}{\partial k \partial X}(k_0,X_0)
  \quad \text{and} \quad
  C_p \coloneq \frac{\partial^2 E}{\partial X^2}(k_0,X_0).
\end{align*}
\Cref{table:os_parameters} lists the calculated parameters for the critical points $1$ to $6$ labeled in \Cref{fig:lower_bands}, where $\omega_p$ is defined in \cref{os_energy_level} below. Note that $A_pC_p > 0$ {and $B_p \approx 0$ for all these critical points}, so that these six critical points are either local minima or local maxima. {Assuming for simplicity that $B_p$ vanishes, the symbol in the RHS of \cref{eq:Morser} is a sum of a function of $\kappa$ and a function of $Y$, so that its Weyl quantization $\left(\kappa \mapsto -i \frac{d}{dx}, Y \mapsto \epsilon x \right)$, is explicit, given by 
 \begin{align*}
H^{\rm eff}_p \coloneq{}& E(p)-\frac{1}{2} A_p\frac{d^2}{dx^2}+\frac{\epsilon^2}{2}C_p x^2.
\end{align*}
}
This is a quantum harmonic oscillator whose $n$-th ($n \in \mathbb{N}$) eigenvalue is
\begin{align}
\label{os_energy_level}
  E_{p,n,\epsilon} \coloneq E(p) +  \epsilon \,\omega_p\bigg(n+\frac{1}{2}\bigg)
  \quad \text{with} \quad
  \omega_p\coloneq \sign(A_p)\sqrt{A_p C_p}.
\end{align}

\begin{table}[htbp]
  \centering
  \sisetup{table-sign-mantissa, round-mode=places, round-precision=2}
  \caption{Energy, second-order derivatives, and oscillator angular frequency at critical points of the first three bands.\label{table:os_parameters}}
  \begin{tabular}{cS[table-format=-2.2]S[table-format=-2.2]S[table-format=-1.2]S[table-format=-4.2]S[table-format=-3.2]}
    \addlinespace
    \toprule
    {$p$} & \multicolumn{1}{c}{{$E(p)$}} & \multicolumn{1}{c}{{$A_p$}} & \multicolumn{1}{c}{{$B_p$}} & \multicolumn{1}{c}{{$C_p$}} & \multicolumn{1}{c}{{$\omega_p$}} \\
    \midrule
         1 &  -14.046 &  -1.435 &  0.00 &  -132.008 &  -13.762 \\
    \addlinespace
         2 &   -7.662 &   2.618 &  0.00 &   210.261 &   23.461 \\
    3 \& 4 &   10.133 & -32.886 &  0.00 & -1692.295 & -235.907 \\
    \addlinespace
     5 \& 6 &  11.287 &  36.267 &  0.00 &  1910.1673 &  263.204 \\
    \bottomrule
  \end{tabular}
\end{table}

The harmonic approximation hence predicts a concentration of states at discrete energy levels at disregistries and momenta near {the energies of the critical points $p$}.
We note that such spectra are sometimes referred to as ``flat bands'' in the physics literature, despite not precisely corresponding to a constant energy band arising from Bloch theory. Similar situations can be seen, e.g., in the incommensurate double-walled carbon nanotube continuum model in~\cite{dwcnt2015} and the tight-binding model for TMDs in~\cite{carr2020}.
We note that the harmonic approximation predicts a collection of single flat bands, not the overlapping pair of flat bands seen in twisted bilayer graphene~\cite{bistritzer2011moire1}.
Then the approximation of the regularized DoS at the energy $E$ near the critical value $E(p)$ is calculated by
\begin{equation}
  \label{os_dos}
 {\nu_{\epsilon,\sigma}(E)} \approx
 \frac{\epsilon}{|\Omega|}\sum_{\substack{p\in P\\|E(p)-E|\ll 1}}\sum_{n \geq 0} \delta_\sigma\Bigg(E-E(p) -  \omega_{p}\epsilon \bigg(n+\frac{1}{2}\bigg)\Bigg).
\end{equation}

\begin{figure}[htbp]
  \centering
   \subfloat[]{\includegraphics[scale=0.5]{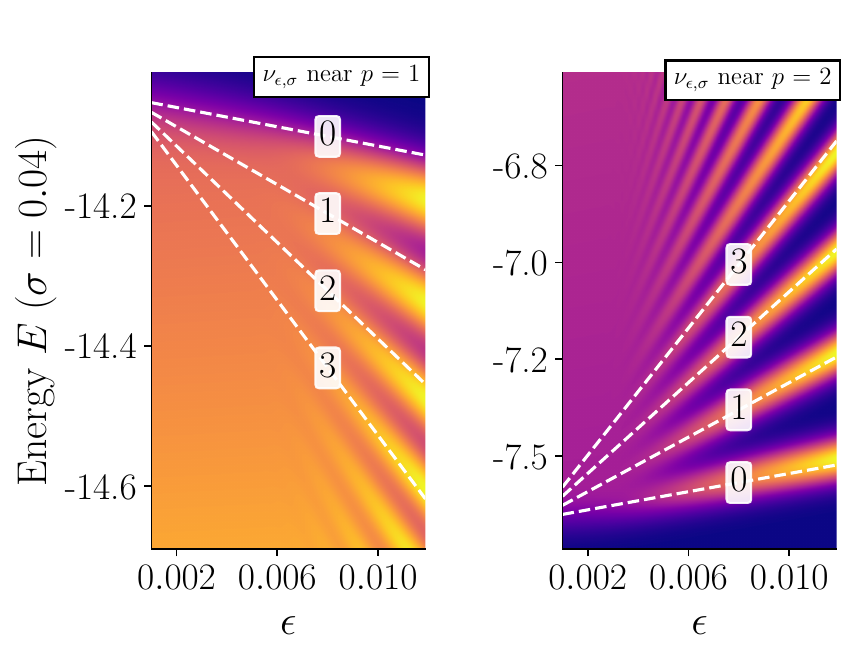}}
    \subfloat[]{\includegraphics[scale=0.5]{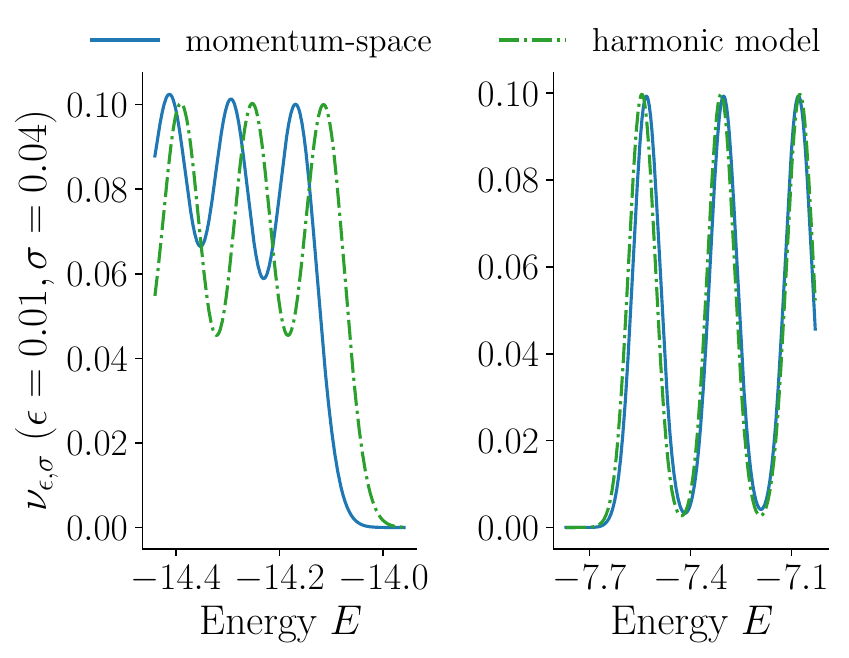}}
  \caption{{Approximations of $\nu_{\epsilon,\sigma}$} in the energy range near $p=1$ and $p=2$ for $\sigma = 0.04$.
    (a)~Comparison of the momentum-space calculation with the harmonic energy levels (dashed white lines) from~\cref{os_energy_level} for $\epsilon$ in $[0.001,0.011]$.
    (b)~Comparison of the momentum-space calculation (solid blue line) with the harmonic approximation from~\cref{os_dos} (dashed green line) for $\epsilon=0.01$. The parameters of the harmonic approximation are listed in \Cref{table:os_parameters}.
    \label{fig:os_blue}}
\end{figure}

\begin{figure}[htbp]
  \centering
    \includegraphics[scale=0.5]{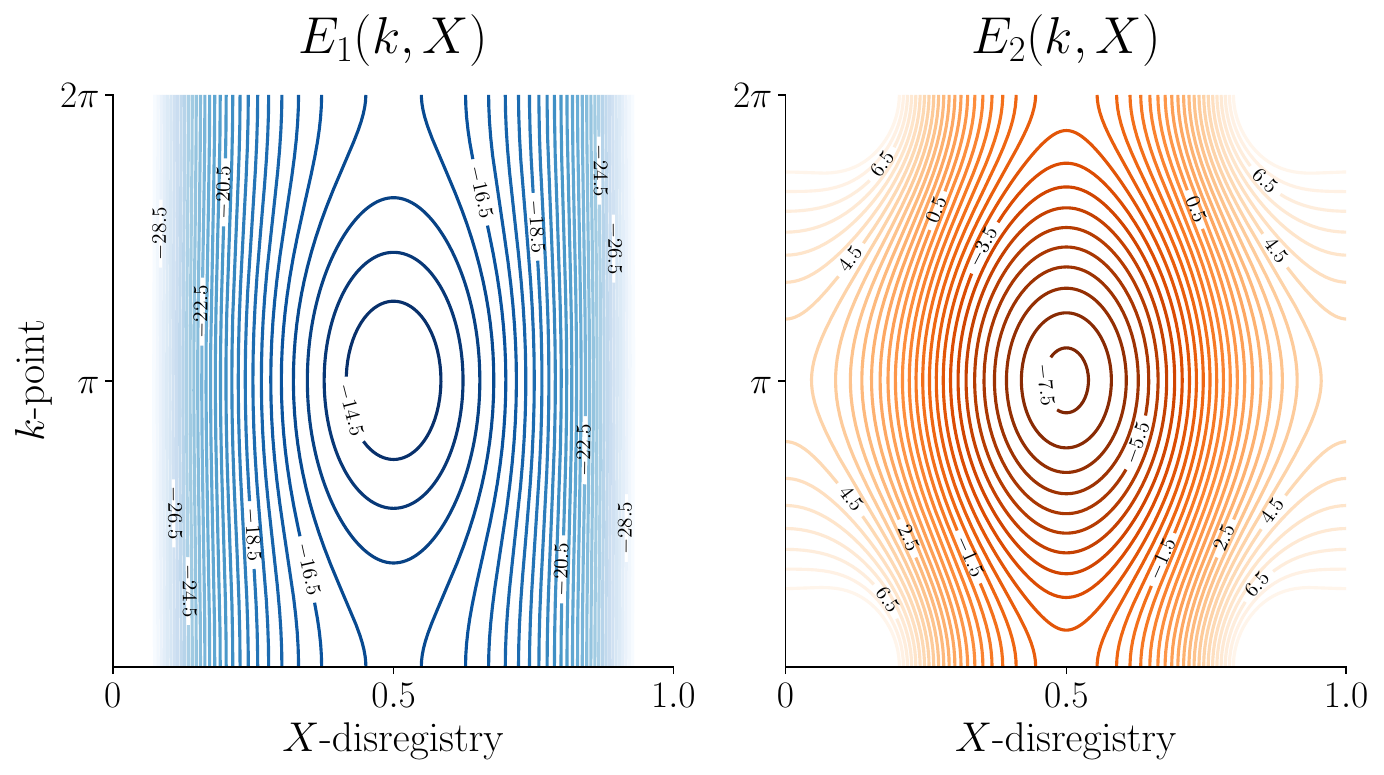}
    \caption{Level sets of $(k,X)\mapsto E_j(k,X)$ for {$j=1$ (Left) and $j=2$ (Right)}.
    \label{fig:contour}}
\end{figure}

\begin{figure}[htbp]
  \centering
    \includegraphics[scale=0.5]{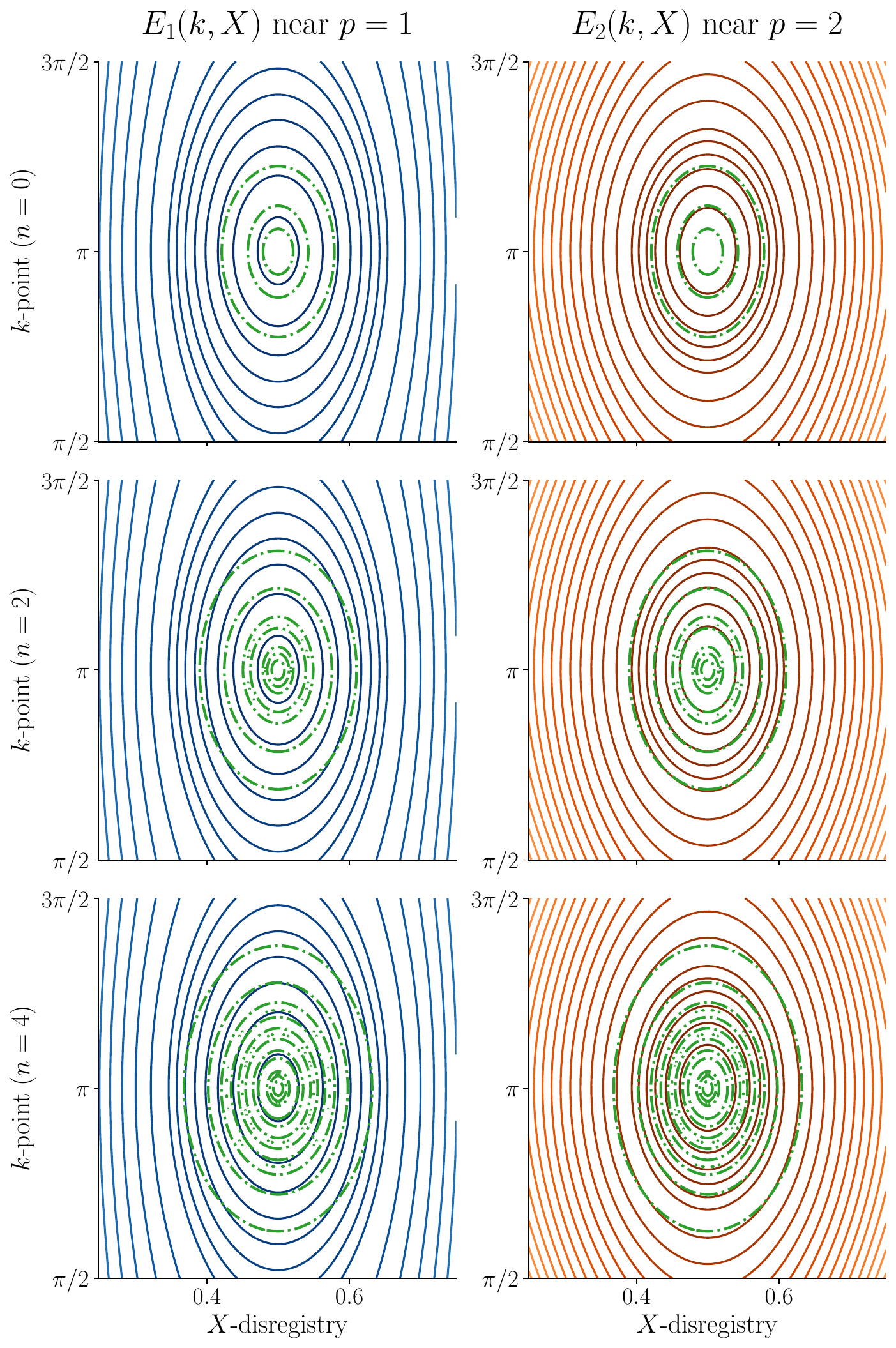}
    \caption{Comparison of the level sets of  $E(k,X)$ with the level sets of $|\mathcal{W}(u_{n,\epsilon})(X,k)|$ (dashed green contours) for $\epsilon=0.01$ and $n=0,2,4$, showing the regions surrounding $p=1$~(Left) and $p=2$~(Right). The parameters of the eigenstates are listed in \Cref{table:os_parameters}.
    \label{fig:contour_zoom}}
\end{figure}

{We first verify the applicability of the harmonic approximation for critical points~1 and~2, both of which are located in relatively flat regions of the band structures. In \Cref{fig:os_blue}, we compare the approximation of the regularized DoS calculated by the momentum-space method~\cref{discrete_pw_dos} for $\sigma=0.04$ with two predictions of the harmonic approximation: (a)~the expected energy levels from~\cref{os_energy_level} for $\epsilon$ in $[0.001,0.011]$, and (b)~the approximation of the regularized DoS given by~\cref{os_dos} for $\epsilon=0.01$.}
We observe excellent agreement between the momentum-space calculation and the harmonic approximation for $p=2$, while the agreement is not as good for $p=1$. 
{This difference can be explained by analyzing the level sets of the functions $(k,X) \mapsto E_j(k,X)$, $j=1,2$, which are plotted in \Cref{fig:contour}. The idea is that if the normalized eigenstates $u_{\epsilon,n}\in L^2(\R)$, $n=0,1,2, \cdots, n_0-1$ of the effective harmonic Hamiltonian around $E(p)$ are localized in the region where the harmonic approximation is valid, the first $n_0$ oscillations should be well-described by $H^{\rm eff}_p$. To quantify this criterion, it is convenient to introduce the Wigner transform of the (shifted and rescaled) eigenstates $u_{\epsilon,n}$, given by 
\begin{align*}
    \mathcal{W}(u_{n,\epsilon})(k,X) \coloneq \frac{1}{\pi \hbar}\int_{\R}u_{\epsilon,n}^*\bigg(\frac{X-X_0}{\epsilon}+y\bigg)u_{\epsilon,n}\bigg(\frac{X-X_0}{\epsilon}-y\bigg)e^{2i\hbar^{-1} (k-k_0) y}\,dy,\qquad \hbar=1.
\end{align*}
Remarkably, the Wigner transform of an eigenfunction of a 1D quantum harmonic oscillator is constant over the level sets of the corresponding classical Hamiltonian (see e.g.,  \cite{mostowski2021,zachos2005quantum}). Plotting the level sets of $|\mathcal{W}(u_{n,\epsilon})(k,X)|$ associated with non-negligible values on top of the level sets of $E(k,X)$, we can expect the harmonic approximation will give satisfactory results if, in the regions of the phase space where $|\mathcal{W}(u_{n,\epsilon})(k,X)|$ is not small, this function takes almost constant values on the level sets of $E(k,X)$. A careful examination of the plots in \Cref{fig:contour_zoom} shows that this is the case for the eigenstates $n=0,2,4$ of $H^{\rm eff}_{p=2}$, but not really for the eigenstates $n=0,2,4$ of $H^{\rm eff}_{p=1}$. 
These observations further explain why, for $\epsilon=0.01$, the harmonic approximation is less accurate for $p=1$ than for $p=2$. 
} 

\begin{figure}[htbp]
  \centering
  \subfloat[]{\includegraphics[scale=0.5]{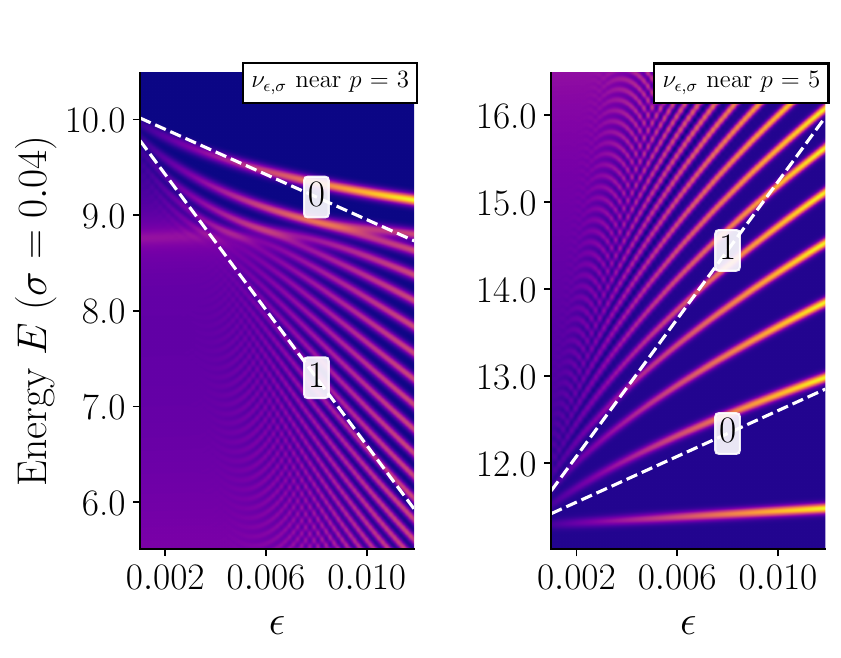}}
    \subfloat[]{\includegraphics[scale=0.5]{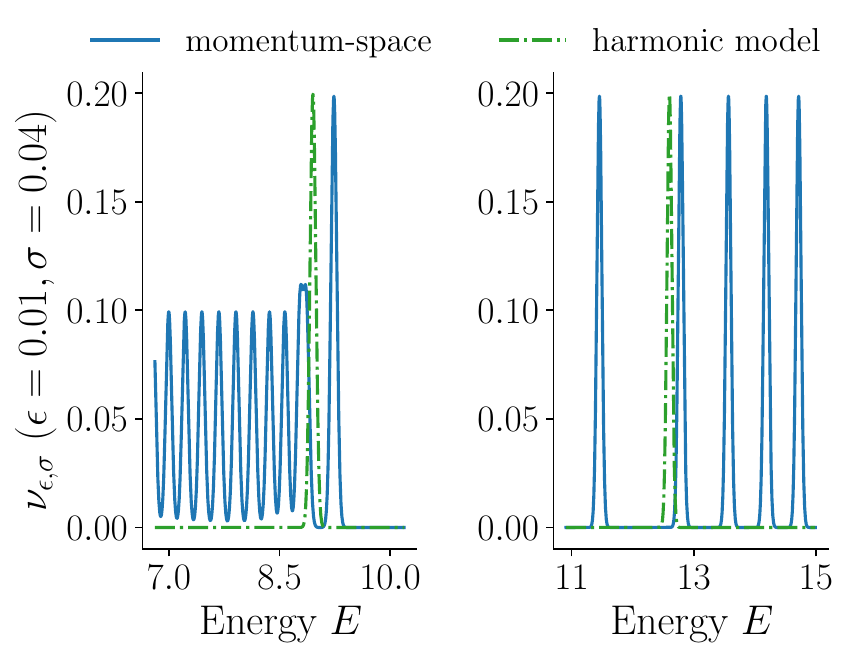}}
  \caption{{Approximations of $\nu_{\epsilon,\sigma}$} in the energy range near $p=3$ and $p=5$ for $\sigma = 0.04$.
    (a)~Comparison of the momentum-space calculation with the harmonic energy levels (dashed white lines) from~\cref{os_energy_level} for $\epsilon$ in $[0.001,0.011]$.
    (b)~Comparison of the momentum-space calculation (solid blue line) with the harmonic approximation from~\cref{os_dos} (dashed green line) for $\epsilon=0.01$. The parameters of the harmonic approximation are listed in \Cref{table:os_parameters}.
    \label{fig:os_orange}}
\end{figure}

We next present results for $p=3$ to $6$ {in the same $\epsilon$ regimes}. Here, we focus on $p=3$ and $p=5$, as the results for $p=4$ and $p=6$ are identical due to the symmetry of the band structures. As shown in \Cref{fig:os_orange}, the harmonic approximation exhibits significant deviations from the momentum-space calculation.
\Cref{fig:lower_bands} shows that the band structures near $p=3$ and $p=5$ are much steeper than those near $p=1$ and $p=2$. {In addition, the local gaps between the second and third bands around $p=3$ and $p=5$ are much smaller than the local gaps between the first and second bands around $p=1$ and $p=2$, leading to non-negligible coupling between bands $2$ and $3$ at the moiré scale at energy around $10$.
While a single band harmonic approximation around $p=1$ and $p=2$ suffices to capture the main features of the DoS in the corresponding energy ranges for the considered values of $\epsilon$, this is not the case for $p=3$ and $p=5$.}

\begin{figure}[htbp]
  \centering
  \subfloat[]{\includegraphics[scale=0.5]{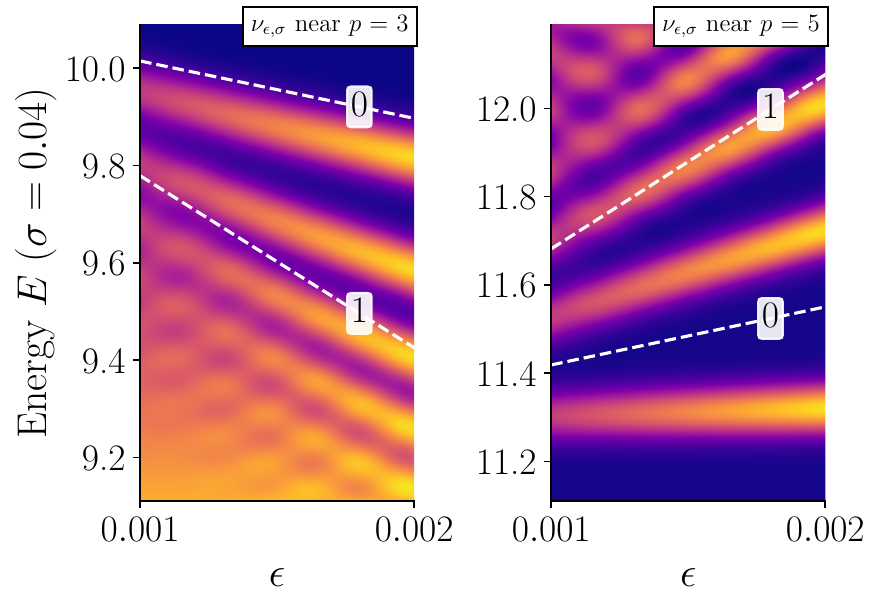}}
    \subfloat[]{\includegraphics[scale=0.5]{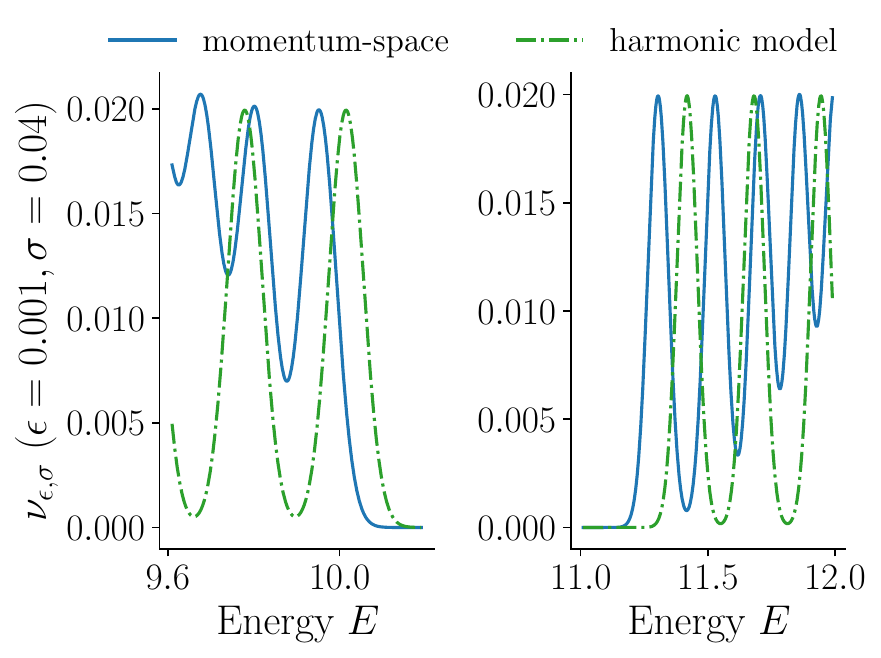}}
  \caption{{Approximations of $\nu_{\epsilon,\sigma}$} in the energy range near $p=3$ and $p=5$ for $\sigma = 0.04$.
    (a)~Comparison of the momentum-space calculation with the harmonic energy levels (dashed white lines) from~\cref{os_energy_level} for $\epsilon$ in $[0.001,0.002]$.
    (b)~Comparison of the momentum-space calculation (solid blue line) with the harmonic approximation~\cref{os_dos} (dashed green line) for $\epsilon=0.001$. The parameters of the harmonic approximation are listed in \Cref{table:os_parameters}.
    \label{fig:os_zoom}}
\end{figure}

{Although the harmonic approximations around $p=3$ to $6$ rapidly deviate from the momentum-space approximation of the regularized DoS as $\epsilon$ increases across the range $[0.001, 0.011]$, we see in \Cref{fig:os_zoom} that the two methods still exhibit good agreement for sufficiently small values of $\epsilon$ within this range}.
From \Cref{table:os_parameters}, we note that the oscillator angular frequency $\omega_p$ is small for $p=1,2$ and large for $p=3$ to $6$. According to the eigenvalue approximation~\cref{os_energy_level}, a large $\omega_p$ causes the eigenvalues $E_{p,n,\epsilon}$ to grow rapidly with $\epsilon$. In addition, note that the effective supports of the Wigner transforms of $u_{n,\epsilon}$ depends on $\epsilon\frac{\omega_p}{A_p}$, although the values of $\frac{\omega_p}{A_p}$ differ only slightly for $p=1$ to 6, a pronounced steepness is observed in the band structures near $p=3$ and $5$. Therefore, the harmonic oscillator model remains valid only for very small values of $\epsilon$ around critical points $p=3,5$.

\section{Conclusion {and perspective}}
In this paper, we have investigated two different methods for approximating the density of states (DoS) of incommensurate Hamiltonians: a momentum-space method (\Cref{sec:PW_method}) and a semiclassical expansion method (\Cref{sec:semiclassical_expansion}). Using a simple 1D toy model and Gaussian smearing, we compared the approximations of the regularized DoS $\nu_{\epsilon,\sigma}$ computed by these two methods.
We first check the consistency of these two methods by comparing the first three terms of the asymptotic expansion in $|\epsilon| \ll 1$ of $\nu_{\epsilon,\sigma}$. Then, we compare the approximations of $\nu_{\epsilon,\sigma}$ for different values of $\epsilon$ and $\sigma$. We observe that, when truncated the semiclassical expansion after second-order, the methods exhibit excellent agreement in the small $\epsilon$ regime, while discrepancies arise for less small $\epsilon$. This indicates the importance of higher-order corrections in the semiclassical method for such regimes. These discrepancies are mainly caused by oscillations in the DoS, which can be analyzed using semiclassical techniques. 

{From a theoretical point of view, both the momentum-space method and the semiclassical expansion method work for higher dimensional incommensurate materials such as twisted bilayer graphene.
This study focuses on a 1D model, which delivers highly accurate results within a computationally feasible timeframe. As such, it provides an ideal framework for demonstrating the two methods used to evaluate the density of states. The insights derived from the 1D case not only validate our approaches but also offer valuable guidance for exploring the DoS in higher-dimensional systems across various regimes—such as different values of $\epsilon$ and the analyticity or smoothness of test functions. Additionally, since Van Hove singularities are also present in higher-dimensional incommensurate systems, the harmonic approximation analysis applied at the band edges in the 1D case can be extended to investigate these phenomena in higher dimensions.
}

{\section*{Data and code availability}
The supporting data and an archived copy of the source codes used to generate the results in this paper are available at 
\url{https://zenodo.org/records/20214148}. 

The MomentumDOS.jl package, which implements the momentum-space method, is available at
\url{https://github.com/xuequan818/MomentumDOS.jl}. The SacerDOS.jl package, which implements the semiclassical method, is available at
\url{https://github.com/epolack/SacerDOS.jl}.
}

\section*{Acknowledgments}
This project has received funding from the Simons Targeted Grant Award No. 896630 and from the European Research Council (ERC) under the European Union's Horizon 2020 research and innovation programme (grant agreement EMC2 No 810367).
{DM’s research was supported by AFRL grant FA9550-24-1-0177.}
{The authors are grateful to the anonymous reviewers for their careful reading of the manuscript and their valuable suggestions, which helped improve the paper.}

\appendix
\section{Derivation of formulae \texorpdfstring{\cref{eq:L1,eq:L2}}{(1) and (2)}}
\label[appendix]{sec:Appendix}

In this section, we prove \Cref{th:2.4} from the semiclassical formulae \cref{eq:f0}-\cref{eq:f2} (derived as in \cite{CancesLong}) and the spectral decomposition \cref{eq:h-decomp}. It is easy to see that \Cref{eq:L0} follows immediately from \cref{eq:f0}. Thus we only focus on the derivation of \cref{eq:L1} and \cref{eq:L2} from \cref{eq:f1} and \cref{eq:f2} respectively.

\subsection{Proof of \texorpdfstring{\cref{eq:L1}}{(1)}}
We first decompose $(z-h)^{-1}\{(z-h),(z-h)^{-1}\}$ by using \cref{eq:h-decomp}. From
\begin{align*}
	\{(z-h),(z-h)^{-1}\}=\nabla_kh\cdot (z-h)^{-1} \nabla_X h(z-h)^{-1} - \nabla_Xh\cdot (z-h)^{-1} \nabla_k h(z-h)^{-1},
\end{align*}
we get
\begin{align*}
	\MoveEqLeft  (z-h)^{-1}\{(z-h),(z-h)^{-1}\}\\
	&=(z-h)^{-1}\nabla_k h\cdot (z-h)^{-1}\nabla_X h (z-h)^{-1}-   (z-h)^{-1}\cdot\nabla_X h(z-h)^{-1}\nabla_k h(z-h)^{-1}.
\end{align*}
Using the fact that $\cK_{mn}=\overline{\cK}_{nm}$ and $\cX_{mn}=\overline{\cX}_{nm}$, we get
\begin{align*}
	\MoveEqLeft   \Tr_{L^2_{\rm per}}[ \{(z-h)^{-1},(z-h)\}(z-h)^{-1}]=2i\sum_{m,n}(\zeta-\lambda_{m})^{-2}(\zeta-\lambda_{n})^{-1} \Im  (\cK_{mn} \cdot \cX_{nm}),
\end{align*}
from which we get
\begin{align*}
	\Tr_{L^2_{\rm per}}[f_{1}]&=-\frac{1}{2}\sum_{m,n} f_2^{(2)}(\lambda_{m};\lambda_{n}) \Im  (\cK_{mn} \cdot \cX_{nm}).
\end{align*}
This gives
\begin{align*}
	L_1(f)=-\frac{1}{2(2\pi)^d}\sum_{m,n}\fint_{\Omega}\int_{\Omega^*}f_2^{(2)}(\lambda_{m}(k,X);\lambda_{n}(k,X)) \Im  (\cK_{mn}(k,X) \cdot \cX_{nm}(k,X)) \,dk\,dX.
\end{align*}

\subsection{Proof of \texorpdfstring{\cref{eq:L2}}{(2)}}
First, we observe that
\begin{align*}
	\MoveEqLeft  (z-h)^{-1}\{(z-h),(z-h)^{-1}\{(z-h),(z-h)^{-1}\}\}\\
	&=(z-h)^{-1}\{(z-h),(z-h)^{-1}\}\{(z-h),(z-h)^{-1}\}\\
	&+(z-h)^{-1}\nabla_X(z-h)(z-h)^{-1}\cdot\nabla_k\{(z-h),(z-h)^{-1}\}\\
	&-(z-h)^{-1}\nabla_k(z-h)(z-h)^{-1}\cdot\nabla_X\{(z-h),(z-h)^{-1}\}\\
	&=(z-h)^{-1}\{(z-h),(z-h)^{-1}\}^2- \{(z-h)^{-1},\{(z-h),(z-h)^{-1}\}\}.
\end{align*}
Thus,
\begin{align*}
	\Tr_{L^2_{\rm per}}[f_2](k,X)=I(k,X)+II(k,X)+III(k,X),
\end{align*}
with
\begin{align*}
	I& \coloneq -\frac{1}{\pi}\int_\C \overline{\partial}\widetilde{f}(z) \Big[-\frac{1}{4} \Tr_{L^2_{\rm per}}[(z-h)^{-1}\{(z-h),(z-h)^{-1}\}^2] \Big]\,dL(z)\\
	II& \coloneq -\frac{1}{\pi}\int_\C \overline{\partial}\widetilde{f}(z)\Big[\frac{1}{4}  \Tr_{L^2_{\rm per}}[\{(z-h)^{-1},\{(z-h),(z-h)^{-1}\}\}]\Big]\,dL(z)\\
	III& \coloneq -\frac{1}{\pi}\int_\C  \overline{\partial}\widetilde{f}(z) \Big[\frac{1}{8} \Tr_{L^2_{\rm per}}[(z-h)^{-1}\{(z-h),(z-h)^{-1}\}_2]\Big]\,dL(z).
\end{align*}

\subsubsection{Reformulation of \texorpdfstring{$I$}{I}}
As
\begin{align*}
	\{(z-h),(z-h)^{-1}\}=\nabla_kh\cdot (z-h)^{-1} \nabla_X h(z-h)^{-1} - \nabla_Xh\cdot (z-h)^{-1} \nabla_k h(z-h)^{-1},
\end{align*}
we have
\begin{align*}
	\MoveEqLeft  (z-h)^{-1}\{(z-h),(z-h)^{-1}\}^2\\
	&=(z-h)^{-1}\Big(\nabla_k h\cdot (z-h)^{-1}\nabla_X h\Big) (z-h)^{-1}\Big(\nabla_kh\cdot (z-h)^{-1} \nabla_X h\Big)(z-h)^{-1}\\
	&\quad+(z-h)^{-1}\Big(\nabla_X h\cdot(z-h)^{-1} \nabla_k h\Big)(z-h)^{-1}\Big(\nabla_X h\cdot(z-h)^{-1} \nabla_kh\Big) (z-h)^{-1}\\
	&\quad -(z-h)^{-1}\Big(\nabla_k h \cdot(z-h)^{-1}\nabla_X h\Big) (z-h)^{-1}\Big(\nabla_X h \cdot(z-h)^{-1} \nabla_k h\Big)(z-h)^{-1}\\
	&\quad -(z-h)^{-1}\Big(\nabla_X h\cdot(z-h)^{-1}\nabla_k h\Big)(z-h)^{-1}\Big(\nabla_k h \cdot(z-h)^{-1}\nabla_X h\Big)(z-h)^{-1}
	\\& \eqcolon I_1(z)+I_2(z)+I_3(z)+I_4(z).
\end{align*}
Using the spectral decomposition~\cref{eq:h-decomp}, we obtain
\begin{align*}
	-\frac{1}{\pi}\int_{\C}\overline{\partial}\widetilde{f}(\zeta)\Tr_{L^2_{\rm per}}[I_1(z)]\,dL(\zeta)&=\frac{1}{4!}\sum_{m,n,p,q}f_4^{(4)}(\lambda_m;\lambda_n,\lambda_p,\lambda_q) (\mathcal{K}_{mn}\cdot\mathcal{X}_{np})(\mathcal{K}_{pq}\cdot\mathcal{X}_{qm}),\\
	-\frac{1}{\pi}\int_{\C}\overline{\partial}\widetilde{f}(\zeta)\Tr_{L^2_{\rm per}}[I_2(z)]\,dL(\zeta)&=\frac{1}{4!}\sum_{m,n,p,q}f_4^{(4)}(\lambda_m;\lambda_n,\lambda_p,\lambda_q) (\mathcal{X}_{mn}\cdot\mathcal{K}_{np})(\mathcal{X}_{pq}\cdot\mathcal{K}_{qm}),\\
	-\frac{1}{\pi}\int_{\C}\overline{\partial}\widetilde{f}(\zeta)\Tr_{L^2_{\rm per}}[I_3(z)]\,dL(\zeta)&=-\frac{1}{4!}\sum_{m,n,p,q}f_4^{(4)}(\lambda_m;\lambda_n,\lambda_p,\lambda_q) (\mathcal{K}_{mn}\cdot\mathcal{X}_{np})(\mathcal{X}_{pq}\cdot\mathcal{K}_{qm}),\\
	-\frac{1}{\pi}\int_{\C}\overline{\partial}\widetilde{f}(\zeta)\Tr_{L^2_{\rm per}}[I_4(z)]\,dL(\zeta)&=-\frac{1}{4!}\sum_{m,n,p,q}f_4^{(4)}(\lambda_m;\lambda_n,\lambda_p,\lambda_q) (\mathcal{X}_{mn}\cdot\mathcal{K}_{np})(\mathcal{K}_{pq}\cdot\mathcal{X}_{qm}).
\end{align*}
By symmetry, we get
\begin{align}
	\label{second term:1}
	\MoveEqLeft I=\frac{1}{96}\sum_{m,n,p,q}f_4^{(4)}(\lambda_m;\lambda_n,\lambda_p,\lambda_q)\Big[-2\Re\Big((\mathcal{K}_{mn}\cdot \mathcal{X}_{np})(\mathcal{K}_{pq}\cdot\mathcal{X}_{qm})\Big)\notag\\
	&\hspace{40.0mm}+(\mathcal{K}_{mn}\cdot\mathcal{X}_{np})(\mathcal{X}_{pq}\cdot\mathcal{K}_{qm})+(\mathcal{X}_{mn}\cdot\mathcal{K}_{np})(\mathcal{K}_{pq}\cdot\mathcal{X}_{qm})\Big].
\end{align}

\subsubsection{Reformulation of \texorpdfstring{$II$}{II}}
We have for $l=1,\dots, d$,
\begin{align*}
	\MoveEqLeft   \partial_{k_l}\{(z-h),(z-h)^{-1}\}\\
	&=\partial_{k_l}\sum_{j}\left(\partial_{k_j}h\partial_{X_j}(z-h)^{-1} - \partial_{X_j}h\partial_{k_j}(z-h)^{-1}\right)\\
	&=(z-h)^{-1} \partial_{X_l}h (z-h)^{-1}- \partial_{X_l} h (z-h)^{-2}\\
	&\quad+ \sum_{j}\partial_{k_j} h (z-h)^{-1}\partial_{k_l}h (z-h)^{-1} \partial_{X_j}h (z-h)^{-1}\\
	&\quad+ \sum_j\partial_{k_j} h (z-h)^{-1} \partial_{X_j}h (z-h)^{-1}\partial_{k_l}h (z-h)^{-1}\\
	&\quad- \sum_j\partial_{X_j} h (z-h)^{-1}\partial_{k_j}h (z-h)^{-1}\partial_{k_l} h (z-h)^{-1}\\
	&\quad- \sum_j\partial_{X_j} h (z-h)^{-1}\partial_{k_l}h (z-h)^{-1}\partial_{k_j} h (z-h)^{-1},
\end{align*}
and
\begin{align*}
	\MoveEqLeft \partial_{X_l}\{(z-h),(z-h)^{-1}\}\\
	&=\partial_{X_l}\sum_{j}\left(\partial_{k_j}h\partial_{X_j}(z-h)^{-1} - \partial_{X_j}h\partial_{k_j}(z-h)^{-1}\right)\\
	&=\sum_j\partial_{k_j} h (z-h)^{-1} \partial_{X_j} h (z-h)^{-1} \partial_{X_l} h (z-h)^{-1}\\
	&\quad+\sum_j\partial_{k_j} h (z-h)^{-1} \partial_{X_l} h (z-h)^{-1} \partial_{X_j} h (z-h)^{-1}\\
	&\quad+\sum_{j}\partial_{k_j} h (z-h)^{-1} \partial_{X_j}\partial_{X_l} h (z-h)^{-1}\\
	&\quad- \sum_j\partial_{X_j}\partial_{X_l} h (z-h)^{-1} \partial_{k_j} h (z-h)^{-1}\\
	&\quad-  \sum_j\partial_{X_j} h (z-h)^{-1}\partial_{X_l} h (z-h)^{-1}\partial_{k_j} h (z-h)^{-1}\\
	& \quad-  \sum_j\partial_{X_j} h (z-h)^{-1}\partial_{k_j} h (z-h)^{-1}\partial_{X_l} h (z-h)^{-1}.
\end{align*}
As a result, we have
\begin{align*}
	\MoveEqLeft  \nabla_X (z-h)^{-1} \cdot \nabla_k\{(z-h),(z-h)^{-1}\}\\
	&=\sum_{l}(z-h)^{-1} \partial_{X_l}h(z-h)^{-2} \partial_{X_l}h (z-h)^{-1}\\
	&\quad-\sum_{l} (z-h)^{-1} \partial_{X_l}h(z-h)^{-1} \partial_{X_l} h (z-h)^{-2}\\
	&\quad+ \sum_{j,l}(z-h)^{-1} \partial_{X_l}h(z-h)^{-1}\partial_{k_j} h (z-h)^{-1}\partial_{k_l}h (z-h)^{-1} \partial_{X_j}h (z-h)^{-1}\\
	&\quad+ \sum_{j,l}(z-h)^{-1} \partial_{X_l}h(z-h)^{-1}\partial_{k_j} h (z-h)^{-1} \partial_{X_j}h (z-h)^{-1}\partial_{k_l}h (z-h)^{-1}\\
	&\quad- \sum_{j,l}(z-h)^{-1} \partial_{X_l}h(z-h)^{-1}\partial_{X_j} h (z-h)^{-1}\partial_{k_j}h (z-h)^{-1}\partial_{k_l} h (z-h)^{-1}\\
	&\quad- \sum_{j,l}(z-h)^{-1} \partial_{X_l}h(z-h)^{-1}\partial_{X_j} h (z-h)^{-1}\partial_{k_l}h (z-h)^{-1}\partial_{k_j} h (z-h)^{-1}\\
	& \eqcolon II_1(z)+II_2(z)+II_3(z)+II_4(z)+II_5(z)+II_6(z),
\end{align*}
and
\begin{align*}
	\MoveEqLeft  -\nabla_k (z-h)^{-1} \cdot \nabla_X\{(z-h),(z-h)^{-1}\}\\
	&=-\sum_{j,l}(z-h)^{-1} \partial_{k_l} h (z-h)^{-1}  \partial_{k_j} h (z-h)^{-1} \partial_{X_j} h (z-h)^{-1} \partial_{X_l} h (z-h)^{-1}\\
	&\quad-\sum_{j,l}(z-h)^{-1} \partial_{k_l} h (z-h)^{-1}  \partial_{k_j} h (z-h)^{-1} \partial_{X_l} h (z-h)^{-1} \partial_{X_j} h (z-h)^{-1}\\
	&\quad-\sum_{j,l}(z-h)^{-1} \partial_{k_l} h (z-h)^{-1}  \partial_{k_j} h (z-h)^{-1} \partial_{X_j}\partial_{X_l} h (z-h)^{-1}\\
	&\quad+ \sum_{j,l}(z-h)^{-1} \partial_{k_l} h (z-h)^{-1} \partial_{X_j}\partial_{X_l} h (z-h)^{-1} \partial_{k_j} h (z-h)^{-1}\\
	&\quad+  \sum_{j,l}(z-h)^{-1} \partial_{k_l} h (z-h)^{-1} \partial_{X_j} h (z-h)^{-1}\partial_{X_l} h (z-h)^{-1}\partial_{k_j} h (z-h)^{-1}\\
	& \quad+  \sum_{j,l}(z-h)^{-1} \partial_{k_l} h (z-h)^{-1} \partial_{X_j} h (z-h)^{-1}\partial_{k_j} h (z-h)^{-1}\partial_{X_l} h (z-h)^{-1}\\
	& \eqcolon II_7(z)+II_8(z)+II_9(z)+II_{10}(z)+II_{11}(z)+II_{12}(z).
\end{align*}
Using again~\cref{eq:h-decomp}, we obtain
\begin{align*}
	-\frac{1}{\pi}\int_{\C}\overline{\partial}\widetilde{f}(\zeta)\Tr_{L^2_{\rm per}}[II_1(z)]\,dL(\zeta)&=\frac{1}{3!} \sum_{m,n}f^{(3)}_3(\lambda_m;\lambda_n,\lambda_n)|\mathcal{X}_{mn}|^2\\
	-\frac{1}{\pi}\int_{\C}\overline{\partial}\widetilde{f}(\zeta)\Tr_{L^2_{\rm per}}[II_2(z)]\,dL(\zeta)&=-\frac{1}{3!}\sum_{m,n}f_3^{(3)}(\lambda_m;\lambda_m,\lambda_n)|\mathcal{X}_{mn}|^2\\
	-\frac{1}{\pi}\int_{\C}\overline{\partial}\widetilde{f}(\zeta)\Tr_{L^2_{\rm per}}[II_3(z)]\,dL(\zeta)&=\frac{1}{4!}\sum_{m,n,p,q} f^{(4)}_{4}(\lambda_m;\lambda_n,\lambda_p,\lambda_q)(\mathcal{X}_{mn}\cdot\mathcal{K}_{pq})(\mathcal{K}_{np}\cdot \mathcal{X}_{qm})\\
	-\frac{1}{\pi}\int_{\C}\overline{\partial}\widetilde{f}(\zeta)\Tr_{L^2_{\rm per}}[II_4(z)]\,dL(\zeta)&=\frac{1}{4!}\sum_{m,n,p,q}f^{(4)}_4(\lambda_m;\lambda_n,\lambda_p,\lambda_q)(\mathcal{X}_{mn}\cdot \mathcal{K}_{qm})(\mathcal{K}_{np}\cdot\mathcal{X}_{pq})\\
	-\frac{1}{\pi}\int_{\C}\overline{\partial}\widetilde{f}(\zeta)\Tr_{L^2_{\rm per}}[II_5(z)]\,dL(\zeta)&=-\frac{1}{4!}\sum_{m,n,p,q}f^{(4)}_4(\lambda_m;\lambda_n,\lambda_p,\lambda_q)(\mathcal{X}_{mn}\cdot \mathcal{K}_{qm})(\mathcal{X}_{np}\cdot \mathcal{K}_{pq})\\
	-\frac{1}{\pi}\int_{\C}\overline{\partial}\widetilde{f}(\zeta)\Tr_{L^2_{\rm per}}[II_6(z)]\,dL(\zeta)&=-\frac{1}{4!}\sum_{m,n,p,q}f^{(4)}_4(\lambda_m;\lambda_n,\lambda_p,\lambda_q)(\mathcal{X}_{mn}\cdot \mathcal{K}_{pq})(\mathcal{X}_{np}\cdot\mathcal{K}_{qm})\\
	-\frac{1}{\pi}\int_{\C}\overline{\partial}\widetilde{f}(\zeta)\Tr_{L^2_{\rm per}}[II_7(z)]\,dL(\zeta)&=-\frac{1}{4!}\sum_{m,n,p,q}f^{(4)}_4(\lambda_m;\lambda_n,\lambda_p,\lambda_q) (\mathcal{K}_{mn}\cdot \mathcal{X}_{qm})(\mathcal{K}_{np}\cdot \mathcal{X}_{pq})\\
	-\frac{1}{\pi}\int_{\C}\overline{\partial}\widetilde{f}(\zeta)\Tr_{L^2_{\rm per}}[II_8(z)]\,dL(\zeta)&=-\frac{1}{4!}\sum_{m,n,p,q}f^{(4)}_4(\lambda_m;\lambda_n,\lambda_p,\lambda_q) (\mathcal{K}_{mn}\cdot \mathcal{X}_{pq})(\mathcal{K}_{np}\cdot \mathcal{X}_{qm})\\
	-\frac{1}{\pi}\int_{\C}\overline{\partial}\widetilde{f}(\zeta)\Tr_{L^2_{\rm per}}[II_9(z)]\,dL(\zeta)&=-\frac{1}{3!}\sum_{m,n,p}f^{(3)}_3(\lambda_m;\lambda_n,\lambda_p) (\mathcal{K}_{mn}\cdot \mathcal{X}^{(2)}_{pm} \cdot \mathcal{K}_{np})\\
	-\frac{1}{\pi}\int_{\C}\overline{\partial}\widetilde{f}(\zeta)\Tr_{L^2_{\rm per}}[II_{10}(z)]\,dL(\zeta)&=\frac{1}{3!}\sum_{m,n,p}f^{(3)}_3(\lambda_m;\lambda_n,\lambda_p) (\mathcal{K}_{mn}\cdot \mathcal{X}^{(2)}_{np} \cdot \mathcal{K}_{pm})\\
	-\frac{1}{\pi}\int_{\C}\overline{\partial}\widetilde{f}(\zeta)\Tr_{L^2_{\rm per}}[II_{11}(z)]\,dL(\zeta)&=\frac{1}{4!}\sum_{m,n,p,q}f^{(4)}_4(\lambda_m;\lambda_n,\lambda_p,\lambda_q) (\mathcal{K}_{mn}\cdot\mathcal{X}_{pq})(\mathcal{X}_{np}\cdot \mathcal{K}_{qm})\\
	-\frac{1}{\pi}\int_{\C}\overline{\partial}\widetilde{f}(\zeta)\Tr_{L^2_{\rm per}}[II_{12}(z)]\,dL(\zeta)&=\frac{1}{4!}\sum_{m,n,p,q}f^{(4)}_4(\lambda_m;\lambda_n,\lambda_p,\lambda_q) (\mathcal{K}_{mn}\cdot \mathcal{X}_{qm}) (\mathcal{X}_{np}\cdot  \mathcal{K}_{pq}).
\end{align*}
Summing up the 12 terms, and regrouping some of them, we obtain
\begin{align}
	\label{second term:2}
	\MoveEqLeft II=\frac{1}{24}\sum_{m,n}\Big(f^{(3)}_3(\lambda_m;\lambda_n,\lambda_n)-f^{(3)}_3(\lambda_m;\lambda_m,\lambda_n)\Big)|\mathcal{X}_{mn}|^2\notag\\
	\nonumber
	&+ \frac{1}{24}\sum_{m,n,p}f^{(3)}_3(\lambda_m;\lambda_n,\lambda_p)\Big[ \mathcal{K}_{mn}\cdot \mathcal{X}^{(2)}_{np} \cdot \mathcal{K}_{pm}-
	\mathcal{K}_{mn}\cdot \mathcal{X}^{(2)}_{pm} \cdot \mathcal{K}_{np}\Big]\\
	\nonumber
	&+\frac{1}{96} \sum_{m,n,p,q}f^{(4)}_4(\lambda_m;\lambda_n,\lambda_p,\lambda_q)\Big[(\mathcal{X}_{mn}\cdot \mathcal{K}_{pq})(\mathcal{K}_{np}\cdot \mathcal{X}_{qm})+(\mathcal{K}_{mn}\cdot \mathcal{X}_{pq})(\mathcal{X}_{np}\cdot\mathcal{K}_{qm})\notag\\
	&\quad+2\Re \big((\mathcal{X}_{mn}\cdot \mathcal{K}_{qm})(\mathcal{K}_{np}\cdot \mathcal{X}_{pq}-\mathcal{X}_{np}\cdot \mathcal{K}_{pq})\big)  -2\Re \big((\mathcal{X}_{mn}\cdot \mathcal{K}_{pq})(\mathcal{X}_{np}\cdot \mathcal{K}_{qm})\big)\Big].
\end{align}

\subsubsection{Reformulation of \texorpdfstring{$III$}{III}}
We recall that
\begin{align*}
	\{a,b\}_2=\sum_{1\leq j,l\leq d}(\partial_{k_j}\partial_{k_l}a \partial_{X_j}\partial_{X_l}b+\partial_{X_j}\partial_{X_l}a\partial_{k_j}\partial_{k_l}b-2\partial_{k_j}\partial_{X_l}a\partial_{X_j}\partial_{k_l}b).
\end{align*}
We have
\begin{align*}
	\MoveEqLeft \{(z-h),(z-h)^{-1}\}_2\\
	&=\sum_{1\leq j,l\leq d}\partial_{k_j}\partial_{k_l} (z-h)\partial_{X_j}\partial_{X_l} (z-h)^{-1}+\sum_{1\leq j,l\leq d}\partial_{X_j}\partial_{X_l} (z-h)\partial_{k_j}\partial_{k_l} (z-h)^{-1}\\
	&\quad -2\sum_{1\leq j,l\leq d}\partial_{k_j}\partial_{X_l} (z-h)\partial_{X_j}\partial_{k_l} (z-h)^{-1}\\
	&=- 2\sum_{j}(z-h)^{-1}\partial_{X_j} h (z-h)^{-1}\partial_{X_j}h (z-h)^{-1} -\sum_{j}(z-h)^{-1}\partial_{X_j}^2 h (z-h)^{-1} \\
	&\quad-2 \sum_{j,l}\partial_{X_j}\partial_{X_l} h (z-h)^{-1}\partial_{k_l} h (z-h)^{-1}\partial_{k_j} h (z-h)^{-1} -\sum_{j}\partial_{X_j}^2 h (z-h)^{-2}.
\end{align*}
As a result,
\begin{align*}
	\MoveEqLeft (z-h)^{-1}\{(z-h),(z-h)^{-1}\}_2\\
	&=\sum_{1\leq j,l\leq d}\partial_{k_j}\partial_{k_l} (z-h)\partial_{X_j}\partial_{X_l} (z-h)0^{-1}+\sum_{1\leq j,l\leq d}\partial_{X_j}\partial_{X_l} (z-h)\partial_{k_j}\partial_{k_l} (z-h)^{-1}\\
	&\quad -2\sum_{1\leq j,l\leq d}\partial_{k_j}\partial_{X_l} (z-h)\partial_{X_j}\partial_{k_l} (z-h)^{-1}\\
	&=- 2\sum_{j}(z-h)^{-2}\partial_{X_j} h (z-h)^{-1}\partial_{X_j}h (z-h)^{-1} -\sum_{j}(z-h)^{-2}\partial_{X_j}^2 h (z-h)^{-1} \\
	&\quad-2 \sum_{j,l}(z-h)^{-1}\partial_{X_j}\partial_{X_l} h (z-h)^{-1}\partial_{k_l} h (z-h)^{-1}\partial_{k_j} h (z-h)^{-1} \\
	&\quad-\sum_{j}(z-h)^{-1}\partial_{X_j}^2 h (z-h)^{-2}\\
	& \eqcolon III_1(z) + III_2(z) + III_3(z) + III_4(z).
\end{align*}
Using~\cref{eq:h-decomp}, we get
\begin{align*}
	-\frac{1}{\pi}\int_{\C}\overline{\partial}\widetilde{f}(\zeta)\Tr_{L^2_{\rm per}}[III_1(z)]\,dL(\zeta)&=-\frac{1}{3}\sum_{m,n}f^{(3)}_3(\lambda_m;\lambda_m,\lambda_n)|\mathcal{X}_{mn}|^2,\\
	-\frac{1}{\pi}\int_{\C}\overline{\partial}\widetilde{f}(\zeta)\Tr_{L^2_{\rm per}}[III_2(z)]\,dL(\zeta)&=-\frac{1}{2}\sum_{m}f''(\lambda_m)\Tr_{\C^d}\mathcal{X}_{mm}^{(2)},\\    -\frac{1}{\pi}\int_{\C}\overline{\partial}\widetilde{f}(\zeta)\Tr_{L^2_{\rm per}}[III_3(z)]\,dL(\zeta)&=-\frac{1}{3}\sum_{m,n,p}f^{(3)}_3(\lambda_m;\lambda_n,\lambda_p)\mathcal{K}_{pm}\cdot \mathcal{X}^{(2)}_{mn}\cdot \mathcal{K}_{np},\\
	-\frac{1}{\pi}\int_{\C}\overline{\partial}\widetilde{f}(\zeta)\Tr_{L^2_{\rm per}}[III_4(z)]\,dL(\zeta)&=-\frac{1}{2}\sum_{m}f''(\lambda_m)\Tr_{\C^d}\mathcal{X}_{mm}^{(2)}.
\end{align*}
Thus, we have
\begin{align}
	\label{second term:3}
	\MoveEqLeft III= -\frac{1}{8}\sum_{m} f''(\lambda_{m})\Tr_{\C^d}\mathcal{X}_{mm}^{(2)}-\frac{1}{24}\sum_{m,n}f^{(3)}_3(\lambda_m;\lambda_m,\lambda_n)|\mathcal{X}_{mn}|^2\notag\\
	&\quad -\frac{1}{24}\sum_{m,n,p}f^{(3)}_3(\lambda_m;\lambda_n,\lambda_p)\mathcal{K}_{pm}\cdot \mathcal{X}^{(2)}_{mn}\cdot \mathcal{K}_{np}.
\end{align}

\subsection{Conclusion}
The second order term is obtained by combining \cref{second term:1}, \cref{second term:2} and \cref{second term:3}:
\begin{align*}
	L_2(f)
	={}&\frac{1}{(2\pi)^d}\fint_{\Omega}\int_{\Omega^*}\Bigg( -\frac{1}{8}\sum_{m} f''(\lambda_{m})\Tr_{\C^d}\mathcal{X}_{mm}^{(2)} \\
	&+ \frac{1}{24}\sum_{m,n}\Big(f^{(3)}_3(\lambda_m;\lambda_n,\lambda_n)-2f^{(3)}_3(\lambda_m;\lambda_m,\lambda_n)\Big)|\mathcal{X}_{mn}|^2\\
	& +\frac{1}{24}\sum_{m,n,p}f^{(3)}_3(\lambda_m;\lambda_n,\lambda_p)\Big[\mathcal{K}_{mn}\cdot \mathcal{X}^{(2)}_{np}\cdot \mathcal{K}_{pm}- 2 \mathcal{K}_{mn}\cdot \mathcal{X}^{(2)}_{pm} \cdot \mathcal{K}_{np}\Big]\\
	&+\frac{1}{96}\sum_{m,n,p,q}f_4^{(4)}(\lambda_m;\lambda_n,\lambda_p,\lambda_q)\Big[(\mathcal{X}_{mn}\cdot\mathcal{K}_{np})(\mathcal{K}_{pq}\cdot\mathcal{X}_{qm})\notag\\
	&\qquad+(\mathcal{X}_{mn}\cdot \mathcal{K}_{pq})(\mathcal{K}_{np}\cdot \mathcal{X}_{qm})+(\mathcal{K}_{mn}\cdot \mathcal{X}_{pq})(\mathcal{X}_{np}\cdot\mathcal{K}_{qm})+(\mathcal{K}_{mn}\cdot\mathcal{X}_{np})(\mathcal{X}_{pq}\cdot\mathcal{K}_{qm})\notag\\
	&\qquad -2\Re\Big((\mathcal{K}_{mn}\cdot \mathcal{X}_{np})(\mathcal{K}_{pq}\cdot\mathcal{X}_{qm})\Big)-2\Re \big((\mathcal{X}_{mn}\cdot \mathcal{K}_{pq})(\mathcal{X}_{np}\cdot \mathcal{K}_{qm})\big)\notag\\
	&\qquad +2\Re \big((\mathcal{X}_{mn}\cdot \mathcal{K}_{qm})(\mathcal{K}_{np}\cdot \mathcal{X}_{pq}-\mathcal{X}_{np}\cdot \mathcal{K}_{pq})\big)\Big]\Bigg)(k,X)\,dk\,dX.
\end{align*}

\subsection{Reduction to \texorpdfstring{$1$}{1}D  case}
For $1$D problems, the above formula can be simplified:
\begin{align*}
	L_2(f) ={}&
	\int_{-1/2}^{1/2} \int_{-\pi}^\pi \bigg(-\frac{1}{8}\sum_{m} f''(\lambda_{m}) \mathcal{X}_{mm}^{(2)} + \frac{1}{24}\sum_{m,n}\Big(f^{(3)}_3(\lambda_m;\lambda_n,\lambda_n)-2f^{(3)}_3(\lambda_m;\lambda_m,\lambda_n)\Big)|\mathcal{X}_{mn}|^2\\
	& +\frac{1}{24}\sum_{m,n,p}f^{(3)}_3(\lambda_m;\lambda_n,\lambda_p)\Big[\mathcal{K}_{mn} \mathcal{X}^{(2)}_{np} \mathcal{K}_{pm}- 2 \mathcal{X}^{(2)}_{mn} \mathcal{K}_{np}  \mathcal{K}_{pm} \Big]\\
	&+\frac{1}{48}\sum_{m,n,p,q}f_4^{(4)}(\lambda_m;\lambda_n,\lambda_p,\lambda_q) \Big[\mathcal{X}_{mn}\mathcal{K}_{np}\mathcal{K}_{pq}\mathcal{X}_{qm}+\mathcal{K}_{mn} \mathcal{X}_{np}  \mathcal{X}_{pq}\mathcal{K}_{qm} \\
	&\qquad-2\Re(\mathcal{X}_{mn}\mathcal{X}_{np} \mathcal{K}_{pq}\mathcal{K}_{qm})\Big] \bigg) \, dk \, dX.
\end{align*}
This ends the proof of \Cref{th:2.4}.

\medskip

\bibliographystyle{siamplain}
\bibliography{references}

\end{document}